\documentclass{jfm}
\renewcommand{\absfooterflag}{}
\renewcommand{\pagelimitfooter}{}
\usepackage{graphicx}
\usepackage{newtxtext}
\usepackage{newtxmath}
\usepackage{natbib}
\usepackage{soul}
\usepackage{amsmath}
\usepackage{amssymb}
\usepackage{subcaption}
\usepackage[export]{adjustbox}
\usepackage{bm}
\usepackage{listings}
\usepackage{xcolor}
\usepackage{hyperref}
\usepackage{booktabs}
\hypersetup{
    colorlinks = true,
    urlcolor   = blue,
    citecolor  = black,
}

\newcommand{\beq}{\begin{equation}}
\newcommand{\eeq}{\end{equation}}
\newcommand{\bee}{\begin{eqnarray}}
\newcommand{\eee}{\end{eqnarray}}
\newcommand{\RomanNumeralCaps}[1]
\lstdefinestyle{pyJFM}{
  language=Python,
  basicstyle=\ttfamily\small,
  keywordstyle=\color{blue!70!black}\bfseries,
  commentstyle=\color{green!40!black},
  stringstyle=\color{red!60!black},
  numberstyle=\scriptsize\color{black!60},
  numbers=left,
  numbersep=8pt,
  stepnumber=1,
  showstringspaces=false,
  breaklines=true,
  breakatwhitespace=true,
  columns=fullflexible,
  keepspaces=true,
  tabsize=4,
  frame=single,
  rulecolor=\color{black!25},
  frameround=tttt,
  captionpos=b,
  xleftmargin=2.2em,
  framexleftmargin=1.8em
}

\title{Exact solution for the motion of a rigid particle with $\bm{S_4}$ and $\bm{C_{2v}}$ symmetry settling under gravity in a viscous fluid}

\author{Piotr Zdybel\aff{1}
  \corresp{\email{pzdybel@ippt.pan.pl}}
 \and Maria L. Ekiel-Je\.zewska\aff{1}\corresp{\email{mekiel@ippt.pan.pl}}}

\affiliation{\aff{1} Division of Complex Fluids, Department of Biosystems and Soft Matter, Institute of Fundamental Technological Research, Polish Academy of Sciences, Pawi\'nskiego 5b, 02-106 Warsaw, Poland}

\begin{document}
\maketitle

\begin{abstract}
We provide an exact and complete solution for the dynamics of a rigid particle of uniform density with $C_{2v}$ and $S_4$ symmetry, settling under gravity in a viscous fluid at a Reynolds number much smaller than unity. 
We assume that the fluid sticks to the body surface, and the fluid flow satisfies the Stokes equations. The $S_4$ symmetry renders the problem exactly integrable, with all orbits labelled by a single conserved quantity $0\le C\le 1$.
We determine exactly the time-dependent: orientation of the particle symmetry axis and centre-of-mass position for the whole range of $C$. We show that, for $0<C<1$, there are two different time scales, what leads to quasi-periodic evolution,
 with a significant time-dependent horizontal displacement.
  We obtain the tilt $\theta$ and spin $\psi$ Euler angles as periodic Jacobi elliptic functions of time, and the azimuthal angle $\phi$ through an incomplete elliptic integral of
the third kind with a complex characteristic, which splits $\phi$  into a uniform drift and a strictly periodic modulation. The orientation period and the drift rate (corresponding to a constant angular velocity around the gravity direction)  follow in a closed form.
The vertical centre-of-mass displacement is obtained in terms of periodic in time incomplete elliptic integrals, and the periodic 
orbit-averaged settling velocity reduces to a single ratio of complete elliptic integrals. The horizontal component of the motion in the laboratory frame of reference  is obtained exactly and algebraically in terms of all three Euler angles, so it is quasi-periodic. The horizontal component of the centre-of-mass position traces rosette-like, in general open curves confined by two concentric `envelope' circles.  
We determine very simple exact expressions for radii of both envelopes and demonstrate that they tend to infinity for a family of shapes with the rotation-translation coupling decreasing to zero. We also explain the origin of cusps at the rosette-like trajectory
and provide a commensurability condition selecting strictly periodic
rosettes. The exact results derived in this paper can be applied for the dynamics of arbitrary particles with $C_{2v}$ and $S_4$ symmetries of their shape and density distribution.
\end{abstract}

\section{Introduction}\label{sec:intro}
The sedimentation of rigid particles in viscous fluids at
Reynolds numbers much smaller than unity is a classical problem
in microhydrodynamics, with applications ranging from colloidal
transport \citep{MakinoDoi2005} and atmospheric fibres
\citep{candelier_torques_2025} to biological micro-objects
such as algae and pollen \citep{Ardekani2017}, artificial microswimmers \cite{rizvi}, and natural
irregular aggregates such as river flocs
\citep{gu2025sedimentation}. 
Model particles of
controlled geometry are used in fundamental studies of
shape-dependent settling
\citep{miara2024dynamics,vaquero,joshi}. Recent progress
in three-dimensional reconstruction of a rigid particle orientation   from the images taken by two synchronized cameras
has further expanded the experimental accessibility of these
systems \citep{miara2023light,flapper2025settling,Shekhar2026JFM}.

In the Stokesian regime, the
instantaneous translational and angular velocities of a rigid
body are linearly related to the applied force and torque
through  $6 \times 6$ mobility tensor whose
structure encodes the particle  
geometry \citep{KimKarrila2005}.  The tensorial framework was
established in the seminal work of
\citet{Brenner1963,Brenner1964}, who showed that for non-symmetric rigid bodies under gravity there exists a coupling resulting in their rotational motion. Therefore, properties of different Stokesian propellers have been extensively investigated, with explicit evaluation of the rotational-translational mobility coefficients that determine how fast is the rotation, see, e.g.,  \cite{Brenner1964,HappelBrenner1983,MakinoDoi2003,
KimKarrila2005}.
The specific form of the $6\times6$ mobility tensor depends  
not only on the particle shape but also on the density distribution and the choice of
reference point \citep{KimKarrila2005}. 
Here, we consider the uniform density distribution and evaluate the mobility tensor with respect to
the particle centre of mass.

A central question in the area of the Stokesian dynamics of particles with complex shapes is to what extent 
the particle symmetry dictates the long-term settling behaviour (\cite{Witten2020ShapedColloidalParticles, Sundberg2026FluidInertia}). 
Certain asymmetric micro-objects can hydrodynamically reorient towards a stable, fixed settling configuration; 
\citet{ekiel2009hydrodynamic}
demonstrated this mechanism for chain-like particles composed
of three touching spheres, and \citet{candelier_torques_2025,Shekhar2026JFM} recently extended the analysis to additional shape
families. On the other hand, certain particles
sustain persistent rotational motion and sediment along
geometrically nontrivial paths: for example, rigid helices, anisotropic helicoids, and chiral helical ribbons undergo rotational-translational coupling that results in  helical-like trajectories \citep{Palusa2018,CollinsHelicoid2021,Huseby2025}. Moreover, rigid rods and disks exhibit a constant lateral drift \citep{Clift1978, HappelBrenner1983, KimKarrila2005}.
The above examples have underscored 
the sensitive dependence of settling trajectories on particle
symmetry. Distinguishing \emph{a priori} which of these outcomes - orientational relaxation, persistent rotation,  
lateral drift or other dynamics - is selected for a given particle shape constitutes the central theoretical challenge.

A major recent advance towards resolving this challenge was
made by \citet{joshi}, who considered the class of rigid
bodies possessing two orthogonal planes of symmetry. In the
language of point-group theory, such bodies belong to the
$C_{2v}$ group, generated by two mirror reflections
$\sigma_{xz}$ and $\sigma_{yz}$; the intersection of these reflection planes (i.e., the principal axis 
$z$) 
defines a twofold rotation axis $C_2$. \citeauthor{joshi} showed that,
for any such body settling in Stokes flow, the orientation
dynamics admits a conserved quantity whose value, together
with the structure of the mobility tensor, uniquely determines
whether the body behaves as a \emph{settler} (approaching a
fixed stable vertical orientation), a \emph{drifter} (approaching a
fixed inclined orientation and sedimenting obliquely), or a
\emph{flutterer} (exhibiting persistent, periodic or
quasiperiodic orientational motion with concomitant complex
spatial centre-of-mass trajectories). For flutterers, the dynamics were shown to depend on two distinct timescales, determined numerically: related to periodic oscillations of inclination angles and rotation around vertical axis. The settler/drifter/flutterer classification was verified numerically using the boundary integral method for a family of designed shapes. 
 
In parallel with these theoretical developments, the
experimental and computational programme of \cite{miara2023light,miara2024dynamics,vaquero,vaquero2025fluttering}
has provided detailed characterisations of specific
particle geometries within the flutterer class. Working with
rigid U-shaped disks - achiral particles possessing $C_{2v}$
symmetry - \citet{miara2024dynamics} demonstrated
experimentally that such disks undergo a periodic sequence of
pitching and rolling motions, causing their centre of mass to
sediment along complex trajectories ranging from quasiperiodic
spirals to helices. Crucially, they showed that the handedness
of the trajectory is determined by the initial orientation,
not by the (non-chiral) particle geometry.
\citet{vaquero}, following \cite{gonzalez}, reduced the rigid-body dynamics with six degrees of freedom to two coupled ordinary
differential equations for the time-dependent roll and pitch angles that determine evolution of the yaw angle and of the particle centre of mass, exploiting also the structure of the
numerically computed resistance matrix. A phase-plane analysis
of the resulting system revealed saddles, centres, separatrices, and
closed orbits in good agreement with experimental observations.
More recently, \citet{vaquero2025fluttering} extended this
analysis to U-shaped disks with a broken symmetry (achieved by
pinching along one axis i.e., deforming a circular disk onto the surface of a cone), demonstrating that the fluttering
behaviour is robust to moderate shape perturbations but
undergoes a transition to drifting when the degree of
asymmetry becomes sufficiently large. In this way, classification of the dynamical modes, performed for the rigid shapes of the $C_{2v}$ symmetry, has been extended for a class of shapes with a lower symmetry obtained by a controlled perturbation of the $C_{2v}$ geometry.

Motivated by these developments, the present paper focuses on
a more restrictive yet analytically tractable subclass of
sedimenting bodies: rigid particles possessing both $C_{2v}$
symmetry (two orthogonal mirror planes) and the
roto-reflection symmetry $S_4$ (rotation by $\pi/2$ about the
principal axis composed with reflection in the plane
orthogonal to it). This additional $S_4$ constraint is the key to exact solvability. The plan of the paper is the following. Sec.~\ref{sec:theory} contains the theoretical description. The exact solution for the time-dependence of the Euler angles, specifying the symmetry axis orientation, is provided in Sec.~\ref{sec:exact_euler}. The motion of the particle centre of mass is determined exactly in Sec.~\ref{sec:cent_mass}.
Sec.~\ref{sec:conclusions} contains the conclusions.

%---------------------------------\\

\section{Theoretical background}\label{sec:theory}
\begin{figure}
    \centering
    \begin{subfigure}{0.49\textwidth}
        \centering
        \includegraphics[width=0.75\textwidth]{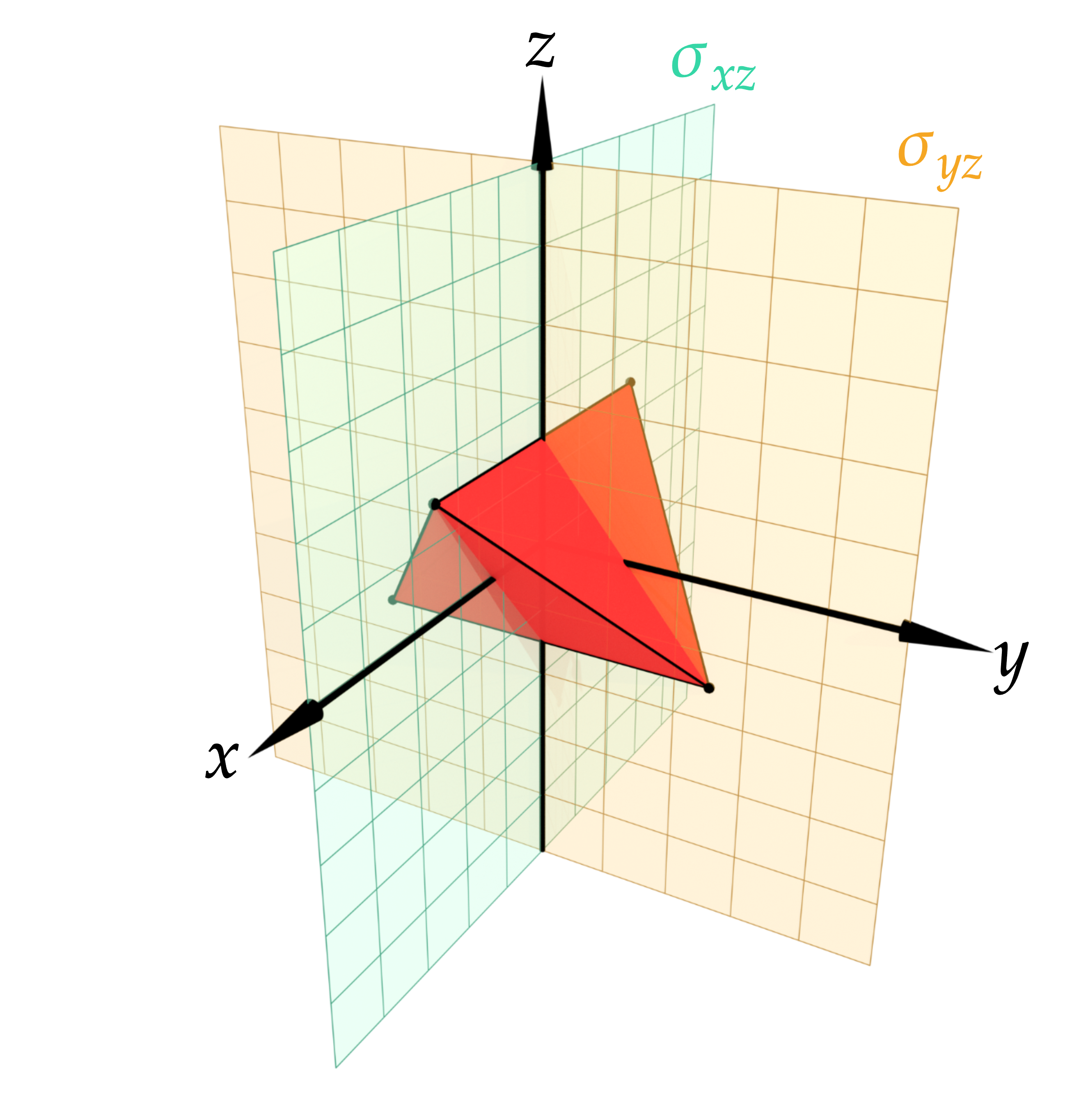}
        \vspace{-0.75cm}\caption{}
        \label{fig:euler_sch}
    \end{subfigure}
    \hfill
    \begin{subfigure}{0.49\textwidth}
        \centering
        \includegraphics[width=0.75\textwidth]{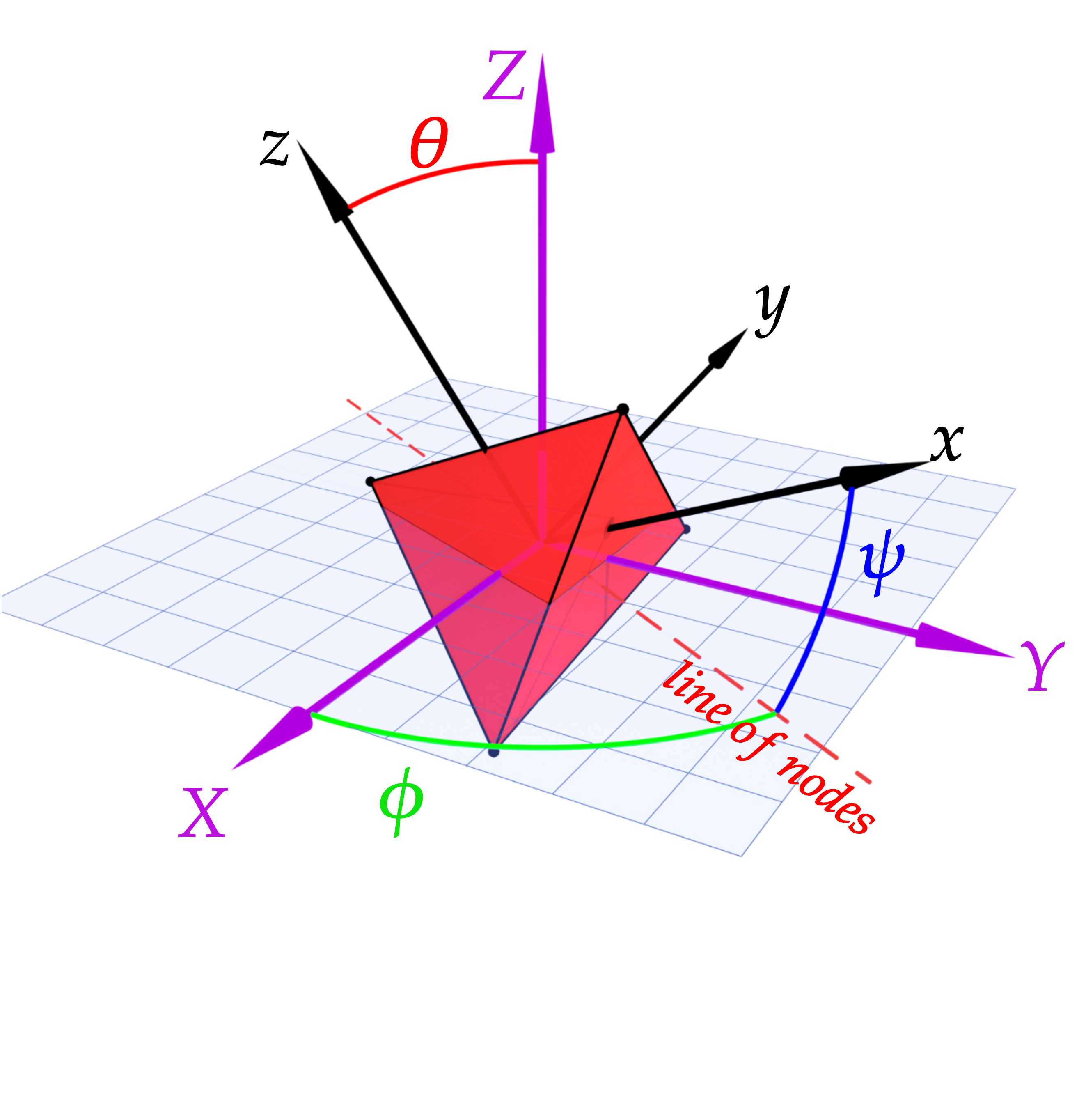}
        \vspace{-0.75cm}\caption{}\label{fig:euler_body_xy}
    \end{subfigure}\vspace{-0.2cm}
    \caption{Shape and parametrization of its orientation.
(a) The exemplary $S_4$- and $C_{2v}$-symmetric particle (elongated tetrahedron). The symmetry planes $\sigma_{xz}\,(y=0)$ and $\sigma_{yz}\,(x=0)$ are indicated (in green and orange, respectively). The $z$-axis coincides with the $S_4$ symmetry axis.
(b)  Orientation of the sedimenting particle expressed by the Euler angles $(\phi,\theta,\psi)$. 
Here, $(x,y,z)$ is the body-fixed frame, whereas $(X,Y,Z)$ are the axes of the laboratory frame, shifted to the particle’s centre of mass.}
\label{euler}
\end{figure}

We focus our attention on a rigid particle with a uniform density distribution. The particle exhibits the rotation-reflection symmetry $S_4$ (the symmetry $C_4=R_z(\pi/2)$ with respect to rotation around $z$ axis by $\pi/2$ superposed with reflection $\sigma_{xy}$  in the $xy$ plane), and the $C_{2v}$ symmetry (superposition of the symmetries $\sigma_{xz}$ and $\sigma_{yz}$ with respect to reflections in $xz$ and $yz$ planes, respectively), see Fig.~\ref{fig:euler_sch}. 

The particle settles under gravity in a viscous fluid with dynamic viscosity~$\eta$.
We assume that the fluid flow satisfies the stationary Stokes equations and that there is no slip of the fluid at the surface of the particle. The dynamics of the chosen shape belong to the (quasi)periodic class of solutions called \textit{flutterers} by \citet{joshi}.

In the symmetric reference frame $xyz$ moving with the particle,  the mobility matrices, evaluated with respect to the particle centre of mass taken as the reference point \citep{KimKarrila2005,Graham2018},
simplify as:
\begin{align}
\mathsfbi{M}^{tt}
=\frac{1}{\pi\eta L}
\begin{pmatrix}
    \mu_1 & 0 & 0\\
    0 & \mu_1 & 0 \\
    0 & 0 & \mu_3
\end{pmatrix}
~~~~~~\mathrm{and}~~~~~
\mathsfbi{M}^{rt}
=\frac{1}{\pi\eta L^2}
\begin{pmatrix}
    0 & \mu_b & 0\\
    \mu_b & 0 & 0 \\
    0 & 0 & 0
\end{pmatrix},
\label{eq:m_tt__m_rr}
\end{align}
where the dimensionless parameters $\mu_1$,  $\mu_3$,  and $\mu_b$ depend on the shape of the rigid particle, and $L$ is a particle length measured along $x$. We assume that $\mu_1\ne\mu_3$, and $\mu_b\ne 0$.
The form of \eqref{eq:m_tt__m_rr} is dictated
entirely by the point group of the particle-shape symmetries, which is generated by the two mirror reflections of
$C_{2v}$ together with the roto-reflection $S_4$. 
Each of these operations maps the particle onto itself, so each of their matrix representations
$\boldsymbol G\in\{\boldsymbol S_4,\boldsymbol\sigma_{xz},\boldsymbol\sigma_{yz}\}$ leaves the
mobility tensors invariant,
\begin{equation}
    \boldsymbol G^{\mathsf T}\mathsfbi M^{tt}\boldsymbol G=\mathsfbi M^{tt},
    \qquad
    \boldsymbol G^{\mathsf T}\mathsfbi M^{rt}\boldsymbol G
      =-\,\mathsfbi M^{rt},
    \label{eq:mobility_invariance}
\end{equation}
with the sign in the second relation reflecting the pseudo-tensor character of the rotational-translational mobility matrix.

The superposition of two mirror reflections
force $\mathsfbi M^{tt}$ to be diagonal and reduce $\mathsfbi M^{rt}$ to its two off-diagonal entries in
the $xy$-block, leaving three independent translational-translational mobility coefficients and two independent rotational-translational mobility coefficients. The additional roto-reflection $S_4$ then identifies the $x$- and $y$-directions and reduces the number of independent mobility coefficients to three: $\mu_1$, $\mu_3$ and $\mu_b$, as shown in equation \eqref{eq:m_tt__m_rr}.

For the centre of mass taken as the reference point, the external torque vanishes, and the dynamics in the laboratory frame of reference have the form,
\begin{align}
\begin{pmatrix}
    \boldsymbol{U}  \\
    \boldsymbol{\omega}
\end{pmatrix}
=
\begin{pmatrix}
    \mathsfbi{M}^{tt}_{lab} \cdot \boldsymbol{F} \\
    \mathsfbi{M}^{rt}_{lab}\cdot \boldsymbol{F} 
\end{pmatrix},
\label{eq:mobility}
\end{align}
where $\boldsymbol{F}$ is the particle's reduced weight (gravity minus buoyancy), $\boldsymbol{U}$ is its centre-of-mass velocity, $\boldsymbol{\omega}$ is its angular velocity, and the mobility matrices in the laboratory reference frame, $\mathsfbi{M}^{tt}_{lab}$ and $\mathsfbi{M}^{rt}_{lab}$, are obtained from equations \eqref{eq:m_tt__m_rr} by the superposition of the corresponding rotations.
We normalize length and time using 
$L$ and 
\beq
\tau_{o}\!=\!\pi\eta L^2/(\mu_b |\boldsymbol{F}|),\label{tauo}
\eeq
respectively. Therefore, $\boldsymbol{U}\!=\!\bm{R}'_{\rm cm}$, with the dimensionless centre-of-mass position \beq\bm{R}_{\rm cm}\!=\!(X_{\rm cm},Y_{\rm cm},Z_{\rm cm}),\label{Rcm}\eeq the dimensionless time~$\tau$, and the time derivative $\mathrm{d}/\mathrm{d}\tau$ denoted by prime '.

We describe the orientational dynamics in \eqref{eq:mobility} using the Euler angles $0<\theta<\pi$, $0\le\psi<2\pi$, and $0\le\phi<2\pi$,  
illustrated in Fig.~\ref{fig:euler_body_xy}, and defined via the line of nodes. 
Let $\boldsymbol{Z}$ be the laboratory axis and $\boldsymbol{z}$ the body-fixed axis; the line of nodes is parallel to the unit vector
\beq
\boldsymbol{n}=\frac{\boldsymbol{Z}\times \boldsymbol{z}}{\|\boldsymbol{Z}\times \boldsymbol{z}\|}\qquad
(\boldsymbol{Z}\not\parallel \boldsymbol{z}).
\eeq
Then $\phi$ is the angle from $\boldsymbol{X}$ to $\boldsymbol{n}$ in the $(\boldsymbol{X},\boldsymbol{Y})$ plane, 
$\theta$ is the angle between $\boldsymbol{Z}$ and $\boldsymbol{z}$, and $\psi$ is the angle from $\boldsymbol{n}$ to $\boldsymbol{x}$ in the plane orthogonal to $\boldsymbol{z}$. 
This corresponds to the rotation matrix
\beq
R(\phi,\theta,\psi)=R_z(\psi)\,R_x(\theta)\,R_z(\phi).
\eeq
The convention of the Euler angles adopted here (with the orientation angles defined by $z$-$x$-$z$ intrinsic
rotations), taken from \citet{Goldstein1980}, follows \citet{ekiel2009hydrodynamic} and \citet{Shekhar2026JFM}. It differs from
the Tait-Bryan parametrisation of the rigid-body orientation by the yaw, pitch and roll angles, used for U-shaped disks by
\citet{miara2024dynamics,vaquero,vaquero2025fluttering}, and care is needed when
comparing phase-portrait coordinates between the two
formulations.

The evolution equations for the Euler angles and the centre-of-mass position read, 
\begin{eqnarray}
\theta' &=&-\cos2\psi\sin \theta, \label{eq:thetapl02}\\
\psi' &=& \sin2\psi \cos \theta \label{eq:psipl02},\\
\phi'&=&-\sin2\psi, \label{eq:phipl02}\\
 \bm{R}'_{\rm cm}&=
& \mathcal{K}(
- \sin 2\theta \sin \phi,\, 
 \sin 2\theta \cos \phi,\,
 -\mathcal{V}_0 +2 \sin^2 \theta), 
\label{eq:}
\end{eqnarray}
with
$\mathcal{K}\!=\!\tfrac{\mu_3 - \mu_1}{2\mu_b}$, and $\mathcal{V}_0\!=\!\frac{2\mu_3}{\mu_3 - \mu_1}$ for $\mu_3\! \ne \!\mu_1$. 
For $\mu_3\! = \!\mu_1$, center of mass moves vertically with a constant speed $-{\mu_3}/{\mu_b}$.  

\section{Exact solution for the Euler angles $\theta$, $\psi$ and $\phi$}
\label{sec:exact_euler}

\subsection{Dynamics of $\theta$ and $\psi$: constant of motion and $\theta$-$\psi$ trajectories}\label{sec:3_1}
As shown by \cite{gonzalez} for an arbitrary shape, the equations for $\theta$ and $\psi$ separate. In case of $C_{2v}$ and $S_4$ symmetries, the first integral of equations \eqref{eq:thetapl02}-\eqref{eq:psipl02} 
 has the form
\begin{equation}
    C=\sin^2\theta\sin2\psi,
    \label{eq:1intl0}
\end{equation}
with $-1 \le C \le 1$, $|C |\le \sin^ 2\theta \le 1$, and $|C| \le |\sin 2\psi| \le 1$. The value of the constant $C$ is specified by the initial conditions as $C=\sin^2\theta_0\sin2\psi_0$, with $\theta_0=\theta(0)$, and $\psi_0=\psi(0)$. The conserved quantity $C$ in~\eqref{eq:1intl0}
is a special case of the invariant identified by \citet{joshi} for the orientational dynamics of the general $C_{2v}$ class of shapes (their Eq.~(5), expressed in a different angular parametrisation).

\begin{figure}
\centerline{\includegraphics[width=0.94\textwidth]{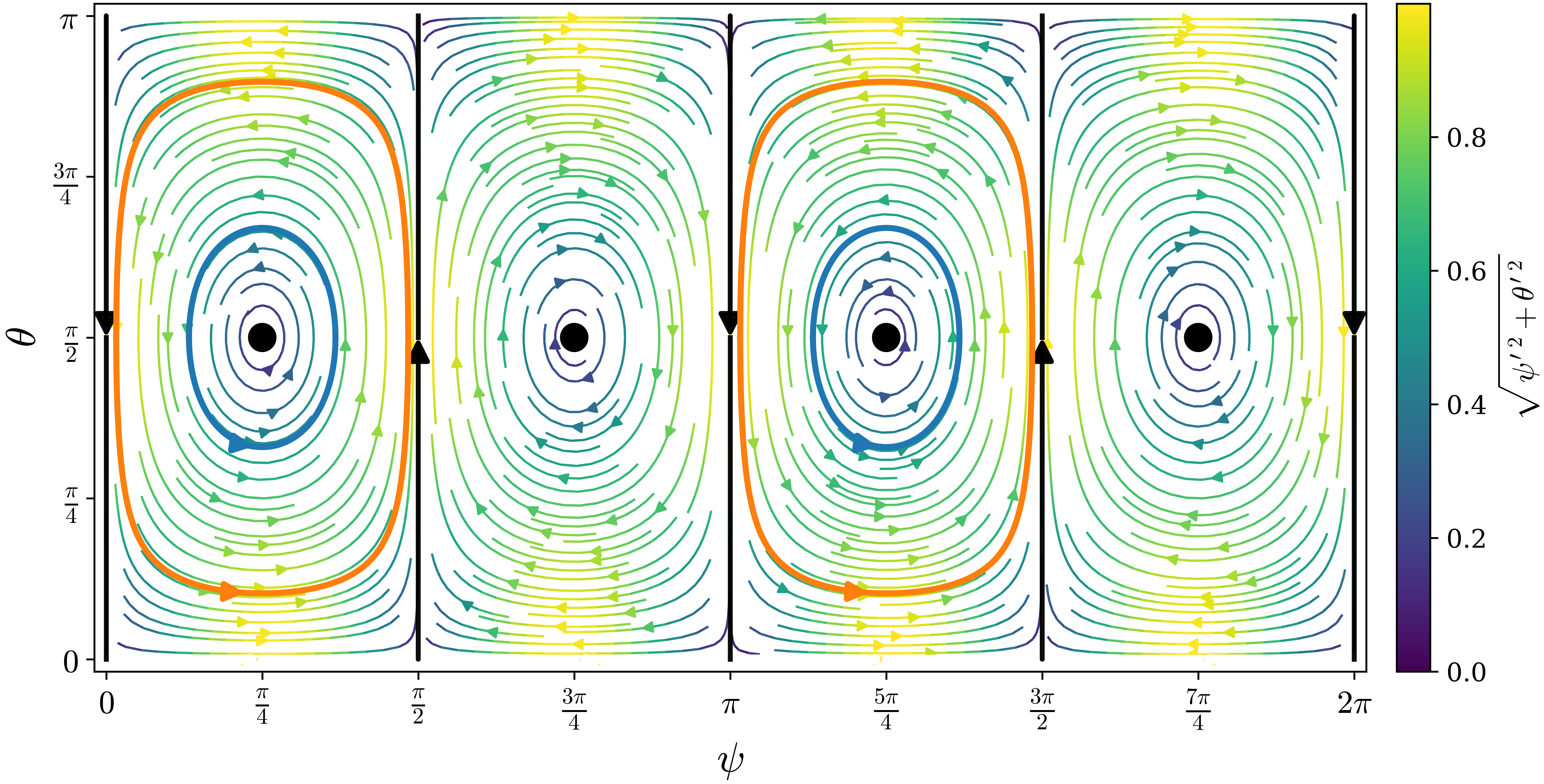}}\vspace{-0.3cm}
\caption{
Phase portrait in the $(\psi,\theta)$ plane for the orientation dynamics of a sedimenting $S_4$- and $C_{2v}$-symmetric body. The orbits with arrows indicating the direction of motion show the vector field $(\psi',\theta')$,  coloured by the phase-space speed $\sqrt{\psi'^{\,2}+\theta'^{\,2}}$. Two representative closed orbits are highlighted for $C=0.10$ (orange) and $C=0.74$ (blue). Fixed points (centres) are marked by black circles at $\theta=\pi/2$ and $\psi =\tfrac{\pi}{4}(2l+1)$, where $l=0, 1, 2, 3$.  The orbits with $\psi (\theta)= 0, \pi/2, \pi, 3\pi/2$ are the separatrices.} 
\label{fig:phase_portrait_S4}
\end{figure}

The trajectories in the $(\psi,\theta)-$space for different values of $C$, following from equation \eqref{eq:1intl0}, are shown in figure \ref{fig:phase_portrait_S4}. The phase-portrait contains centres at $\theta = \pi/2$ surrounded by closed orbits, with separatrices at $\psi = k\pi/2$. The sense of the motion along the trajectories, indicated in figure \ref{fig:phase_portrait_S4}, follows from equations \eqref{eq:thetapl02}-\eqref{eq:psipl02}.
An analogous structure of the phase portrait   
was
obtained numerically by \citet{vaquero} in their figure 5 for
U-shaped disks using Tait--Bryan angles
$(\chi_\mathrm{pitch}, \chi_\mathrm{roll})$.

However, the additional $S_4$ symmetry of the particle shape implies additional symmetry of the trajectory. From equation \eqref{eq:1intl0} it follows that the $\theta$-$\psi$ trajectory is invariant with respect to each of the transformations: 
$(\theta,\psi) \rightarrow (\pi-\theta,\psi)$ and $(\theta,\psi) \rightarrow (\theta,\pi/2-\psi)$. Moreover, \eqref{eq:1intl0} also implies that for $0 < \psi(0) < \pi/2$, the constant of motion $C>0$, and the trajectory stays in the fundamental pocket $0 < \psi(\tau) < \pi/2$. For  $\pi/2 < \psi(0) < \pi$, the constant of motion $C<0$, and the trajectory shape is identical, with the opposite sense of the motion along it. Therefore, it is sufficient to find the solution in the fundamental pocket.

\subsection{Special solutions and general symmetries}

\emph{Special cases.} 
For $C=0$, special solutions of Eqs. \eqref{eq:thetapl02}-\eqref{eq:phipl02} are, as shown by \cite{ekiel2009hydrodynamic}:
\begin{eqnarray}
 \psi(\tau)&=&k\pi/2, \mbox{ with } k=0,1,2,3,\\
\tan(\theta(\tau)/2) &=&
\left\{ 
\begin{array}{l}
A \exp (-\tau) \;\; \mbox{ for } 
k=0,2,\\
A \exp (+\tau) \;\; \mbox{ for } 
k=1,3,
\end{array} 
\right. \label{theta_exp}
\\
\phi(\tau) &=& \phi_0, \label{phi_const}
\end{eqnarray}
with $A=\tan(\theta_0/2)$ and $\phi_0=\phi(0)$.

For $C=\pm 1$ the solution to equations \eqref{eq:thetapl02}-\eqref{eq:psipl02} is the centre, and equation \eqref{eq:phipl02} determines the particle rotation around vertical axis $z$ with a constant angular velocity $\mp 1$,
\bee
\psi(\tau)\!=\!\left\{ \begin{array}{l} \pi/4,\,5\pi/4\;\;\, \mbox{ for }C\!=\!1, \\ 3\pi/4,\,7\pi/4 \;\mbox{ for }C\!=\!-1, \end{array}\right. \hspace{0.5cm} \theta(\tau)\!=\!\pi/2, \hspace{0.5cm} \phi(\tau)\!=\!\left\{ \begin{array}{l} - \tau \!+\! \phi_0\;\;\, \mbox{ for }C\!=\!1,\\ \,+ \tau \!+\! \phi_0\;\;\, \mbox{ for }C\!=\!-1.\end{array}\right.\; 
\label{C=1}
\eee

\emph{General case.} 
We now move to discussing the general case with $0<C<1$. The sign of $C$ equals to the sign of $\sin 2\psi$, which in turn determines the sign of $\phi'$ by Eq.~\eqref{eq:phipl02}. Therefore, the initial value of the angle $\psi$ uniquely  determines the handedness of the helical sedimentation trajectory. This property is valid also for all the particle shapes with the $C_{2v}$ symmetry. It corresponds to the observation of \cite{miara2024dynamics} that the chirality of the trajectory is determined
by the initial orientation rather than by the particle
geometry.

From the right hand side of equations \eqref{eq:thetapl02}-\eqref{eq:phipl02} and the constant of motion \eqref{eq:1intl0} it follows that $\theta'$ and $\psi'$ 
are periodic functions of $\tau$ with a period $T$ (to be evaluated later). In addition, $\phi'$ and $Z'_{cm}$ are periodic functions of $\tau$ with the period $T/2$.  Because of symmetries, $\int_0^T \theta' \mathrm{d}\tau = \int_0^T \psi' \mathrm{d}\tau =0$, and therefore $\theta$ and $\psi$ are also $T$-periodic functions of $\tau$. However, the particle rotates around a vertical axis with a  non-vanishing average angular velocity, $\frac{1}{T}\int_0^T \phi' \mathrm{d}\tau =\Delta \phi /T \ne 0$, and therefore
\beq \phi(\tau)=\tau \Delta \phi/T+\phi_{\rm per}(\tau),\label{phiphiperp}\eeq where $\Delta \phi$ will be determined later, and $\phi_{\rm per}(\tau)$ is periodic with the period $T/2$,
\bee
&&\phi_{\rm per}(\tau+T/2)=\phi_{\rm per}(\tau),\label{phi_per}\\
&&\phi(\tau+T/2)=\phi(\tau)+\Delta \phi/2.\label{phi_half_T} 
\eee
The sign of $\Delta \phi$ depends on the initial value of $\psi$. The characteristic time of the average rotation is $\tau_{\rm drift}=2\pi T/|\Delta \phi|$. 
The subscript ``drift'' is used to indicate the linear increase of $\phi-\phi_{\rm per}$ with time as the result of rotation with the constant angular velocity $\omega_Z=\Delta \phi/T$. 
Because of this linear growth, we find it convenient to extend the range of values of $\phi(\tau)$ beyond $[0,2\pi)$. We will now discuss symmetries of the time-dependent Euler angles.

\emph{Half-period symmetry.}  
The right-hand sides of \eqref{eq:thetapl02}-\eqref{eq:psipl02} are invariant under the point reflection of the $\theta$-$\psi$ trajectory with respect to $(\pi/2,\pi/4)$, 
\begin{equation}
(\theta(\tau),\psi(\tau))\ \longmapsto\ \Bigl(\pi-\theta(\tau),\ \tfrac{\pi}{2}-\psi(\tau)\Bigr),
\label{eq:halfperiod-map}
\end{equation}
as follows from $\sin(\pi-\theta)=\sin\theta$, $\cos(\pi-\theta)=-\cos\theta$, $\cos(\pi-2\psi)=-\cos 2\psi$ and $\sin(\pi-2\psi)=\sin 2\psi$. Consequently, whenever $\bigl(\theta(\tau),\psi(\tau)\bigr)$ solves
\eqref{eq:thetapl02}--\eqref{eq:psipl02}, so does its image under \eqref{eq:halfperiod-map}. The transformed solution traverses the same orbit as the original one, with the reversed signs and the same absolute values of $(\theta'(\tau),\psi'(\tau))$ for every $\tau$. Therefore the times to move from $(\theta(\tau),\psi(\tau))$ to $\Bigl(\pi-\theta(\tau),\ \tfrac{\pi}{2}-\psi(\tau)\Bigr)$ and from $\Bigl(\pi-\theta(\tau),\ \tfrac{\pi}{2}-\psi(\tau)\Bigr)$ to $(\theta(\tau),\psi(\tau))$ are the same, and equal to $T/2$. Therefore,  every closed orbit obeys the half-period relation,
\begin{equation}
\theta\Bigl(\tau+\frac{T}{2}\Bigr)=\pi-\theta(\tau),
\qquad
\psi\Bigl(\tau+\frac{T}{2}\Bigr)=\frac{\pi}{2}-\psi(\tau).
\label{eq:halfperiod}
\end{equation}

\emph{Time reversal.} Every $\theta$-$\psi$ orbit with $0<C<1$ crosses $\theta=\pi/2$, so the time origin $\tau=0$ may always be placed at such an instant that $\theta(0)=\pi/2$. This is a choice of the  the initial position and not a restriction on the orbit. Equations \eqref{eq:thetapl02}--\eqref{eq:psipl02} form a closed autonomous system for $(\theta,\psi)$ alone, and the following transformation leads to another point at the $\theta$-$\psi$ trajectory, reached at another time $\tilde{\tau}$, 
\begin{equation}
(\theta(\tau),\psi(\tau))\ \longmapsto\ (\theta(\tilde{\tau}),\psi(\tilde{\tau}) \equiv \bigl(\ \pi-\theta(\tau),\ \psi(\tau)\bigr),
\label{eq:reversal-map}
\end{equation}
This transformation changes the sign of the right hand side of \eqref{eq:psipl02}; the sign of \eqref{eq:thetapl02} are the same. The absolute values of $\theta'(\tau)$ and $\psi'(\tau)$ remain unchanged. As a consequence, the times to travel from $(\theta(\tau),\psi(\tau))$ to $\bigl(\pi-\theta(\tau),\ \psi(\tau)\bigr)$ and from $\bigl(\pi-\theta(\tau),\ \psi(\tau)\bigr)$ to $(\theta(\tau),\psi(\tau))$ are the same, and $\tilde{\tau}=-\tau$. Therefore,
\begin{equation}
\theta(-\tau)=\pi-\theta(\tau), \qquad \psi(-\tau)=\psi(\tau).
\label{time_reversal}
\end{equation}
The azimuth then follows from \eqref{eq:phipl02} by a single quadrature,
$\phi(\tau)=\phi(0)-\int_{0}^{\tau}\sin 2\psi(\tau')\,\mathrm{d}\tau'$,
whose integrand involves $\psi$ alone. By \eqref{time_reversal} that
integrand is an even function of $\tau$, hence the integral is odd, and the azimuth is reflected about its value at the chosen origin,
\begin{equation}
\phi(-\tau)=2\phi(0)-\phi(\tau).
\label{eq:reversal-phi}
\end{equation}
Subtracting the linear drift $\tau\Delta\phi/T$ of \eqref{phiphiperp}, which is odd in $\tau$, and using $\phi(0)=\phi_{\rm per}(0)$, gives finally
\begin{equation}
\phi_{\rm per}(-\tau)=2\phi_{\rm per}(0)-\phi_{\rm per}(\tau).
\label{phi-}
\end{equation}

\subsection{Reduction to an elliptic equation and Jacobi solution for $\theta(\tau)$ and $\psi(\tau)$}
Consider $0 < C < 1$.
The main task is to solve Eq.\eqref{eq:psipl02}, using the relation \eqref{eq:1intl0}. We obtain 
\begin{equation}
    \left(\psi'\right)^2=\sin^22\psi - C\sin2\psi.
\end{equation} 
We can reduce the problem to an elliptic equation by introducing a new variable
\begin{equation}
    u(\tau):=\tan \psi(\tau) \in  
    [u_-,u_+],
\end{equation}
where the limiting values of $u(\tau)$ are turning points, determined from equation \eqref{eq:1intl0}, 
\beq
u_\pm =\tfrac{1}{C}\left(1\pm \sqrt{1-C^2}\right).\label{pm}
\eeq

This change of variables, with the use of equations \eqref{eq:thetapl02}-\eqref{eq:psipl02}, 
leads to
\begin{equation}
(u')^2  :=-2Cu^3+4u^2-2Cu \;=2C(u-u_-)(u_+-u)u,
    \label{eq:el}
\end{equation}
which is the elliptic differential equation, whose solutions are expressed by Jacobi's elliptic functions. 

We seek a solution of equation \eqref{eq:el} in terms of the Jacobi elliptic function $\mathrm{cn}$, whose
properties are collected in \citet{WhittakerWatson1927,Chandrasekharan1985}; it satisfies the differential equation
\beq
\left(\frac{\mathrm{d}\,\mathrm{cn}(\zeta,k)}{\mathrm{d} \zeta}\right)^2=(1-\mathrm{cn}^2)[1-k^2(1-\mathrm{cn}^2)],
\eeq
with the initial condition $\mathrm{cn}(0,k)=1$ and $0< k< 1$. It follows that $\frac{\mathrm{d}\,\mathrm{cn}(\zeta,k)}{\mathrm{d}\zeta}|_{\zeta=0}=0$. Moreover, $-1 \le\mathrm{cn}(\zeta,k)\le 1$, 
and $\mathrm{cn}(\zeta,k)$ is an even periodic function of $\zeta$ with the period $4K(k^2)$,
where $K(k^2)$ is the complete elliptic integral of the first kind given by
\begin{equation}
    K(k^2)=\int_0^{\pi/2}\frac{\mathrm{d} \vartheta}{\sqrt{1-k^2\sin^2\vartheta}}= \frac{\pi}{2} \;\phantom{}_2F_1\left(\frac{1}{2}, \frac{1}{2};  1; k^2\right)=\frac{\pi}{2}\sum_{n=0}^\infty\left(\frac{(2n-1)!!}{(2n)!!}\right)^2 k^{2n},\label{defK}
\end{equation}
and $\phantom{}_2F_1(a, b; c; x)$ is the hypergeometric function (\cite{Byrd1954Handbook}). 
For $k\rightarrow 0$, $\mathrm{cn}(\zeta,k)\rightarrow \cos \zeta.$ For a comprehensive modern account of the identities and
asymptotic expansions of the Jacobi elliptic functions and
complete elliptic integrals used throughout this work, we
refer the reader to the NIST Digital Library of
Mathematical Functions \citep{DLMF2020}, specifically
chapters~19 (elliptic integrals) and~22 (Jacobi elliptic
functions).\\
\indent We find the solution of equation \eqref{eq:el} as
\begin{equation}
    u(\tau)=u_-+\delta\; \mathrm{cn}^2(\zeta, k), 
    \label{sol_eq}
\end{equation}
where  $\tau_0$ is determined by the relation $u(\tau_0)=u_+$, and
\begin{equation}
 \zeta\!=\!\Omega(\tau\!-\!\tau_0), \quad \delta\!=\!u_+\!-\!u_-\!=\!\frac{2\sqrt{1\!-\!C^2}}{C}, \quad   \Omega^2\!=\!
    \frac{1+\sqrt{1\!-\!C^2}}{2}, 
    \quad k^2
    \!=\!\frac{2\sqrt{1\!-\!C^2}}{1+\sqrt{1\!-\!C^2}}.\label{defi}
\end{equation}

\begin{figure}
    \centering
    \includegraphics[width=0.75\linewidth]{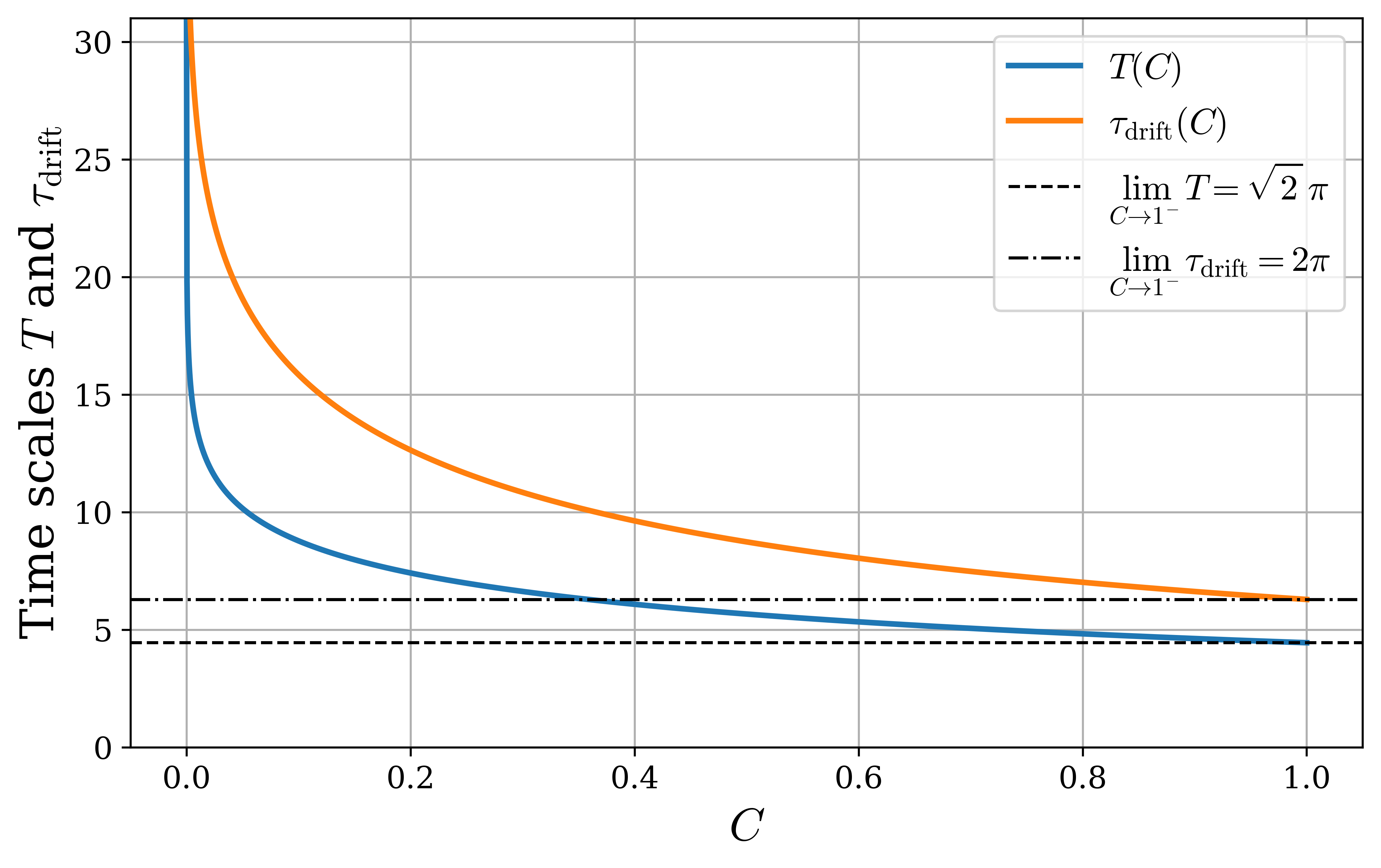}
    \vspace{-0.4cm}
    \caption{
Characteristic time scales $T$ and $\tau_{\rm drift}$ specified by closed forms in Eqs. \eqref{period} and \eqref{eq:tau_drift}, as functions of the first integral $C$ (solid curves). The horizontal dashed and dash-dotted lines indicate the limiting values for ${C\!\to\! 1^-}$. The time unit is defined in Eq.~\eqref{tauo}. 
    }    \label{fig:TvsC}
\end{figure}
The obtained solution $u(\tau)$ is periodic in $\tau$ with the period
\begin{equation}
    T=\frac{2 K(k^2)}{\Omega}, 
    \label{period}
\end{equation}
because the period of $\mathrm{cn}^2(\zeta,k)$ equals to $2K(k^2)$. Therefore, $T$ is also the period of $\psi(\tau)=\text{arctan}\, u(\tau)$. The dependence of the period $T$ on the invariant $C$ is shown in figure \ref{fig:TvsC}.

\begin{figure}
    \centering
    \includegraphics[width=0.85\linewidth]{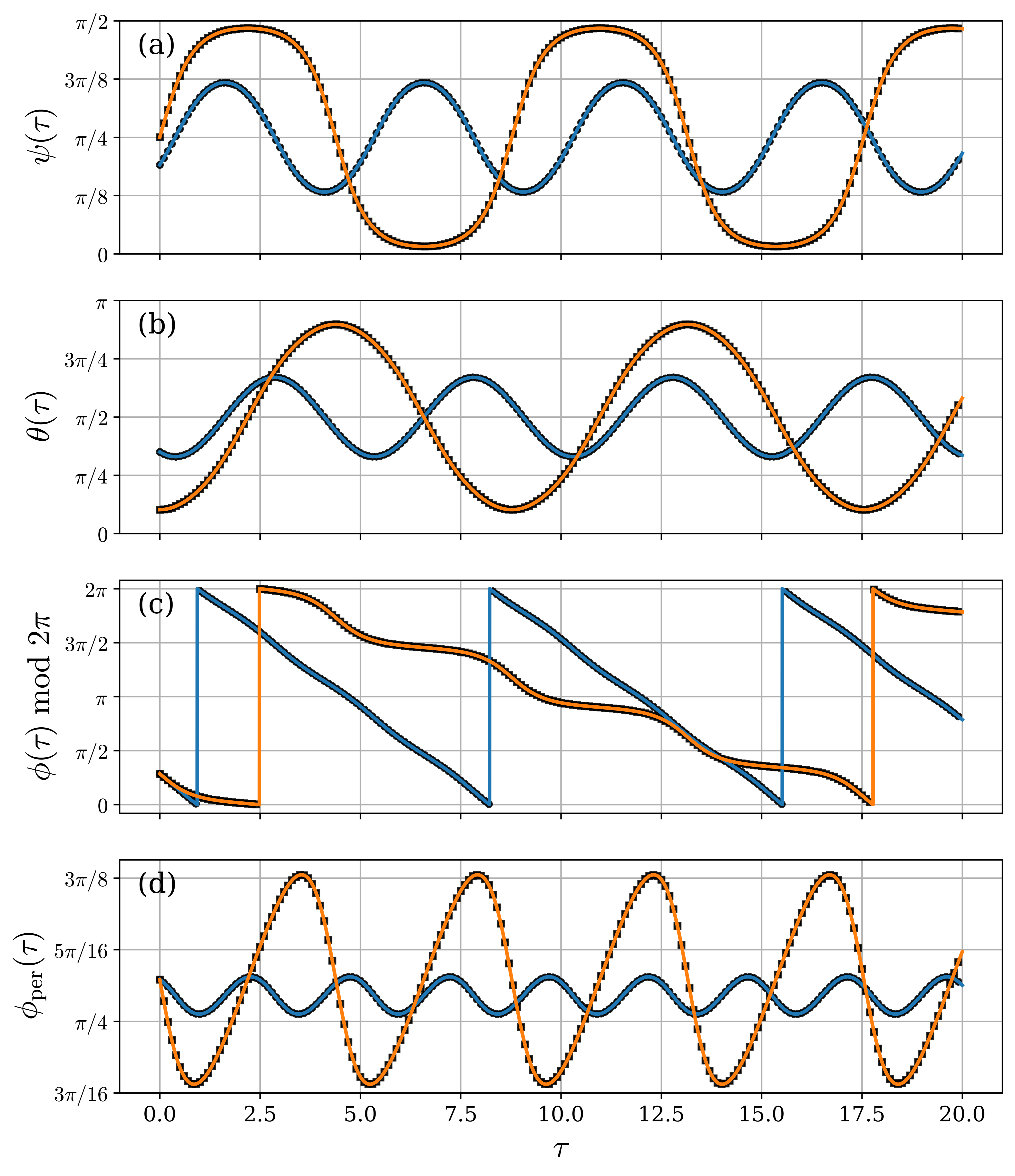}
    \vspace{-0.4cm}
\caption{Exact (closed-form) solutions for the Euler angles (solid colour curves) for two values of the first integral: $C\!=\!0.10$ (orange) and $C\!=\!0.74$ (blue). Panels (a), (b), (c), and (d) show, respectively:  $\psi(\tau)$, $\theta(\tau)$, $\phi(\tau)$ modulo $2\pi$, and   $\phi_{\mathrm{per}}(\tau)=\phi(\tau)\!+\!2\pi \tau/\tau_{\rm drift}$, i.e.,  the azimuthal angle with the subtracted average rotation around $Z$, where  
 $\phi(\tau)$ and $\tau_{\rm drift}$  are evaluated from the expression~\eqref{eq:phi_sol} involving the elliptic integral $\Pi$. Black markers indicate high-accuracy numerical integration of Eqs.~\eqref{eq:thetapl02}-\eqref{eq:phipl02}.}
    \label{fig:angles}
\end{figure}

For $0 < C < 1$,  the solution for $\cos^2 \theta(\tau)$ follows from equation \eqref{eq:1intl0}, 
\beq
\cos^2\theta = 1-\frac{C(u^2+1)}{2u},
\label{theta_u}
\eeq
The function $\theta(\tau)$
is periodic with the same period $T$  given by equation \eqref{period}.
Equation \eqref{theta_u} fixes $\cos^2\theta$, but not the sign of $\cos \theta$, i.e., not the proper half of the domain $(0,\pi)$ for $\theta$. For each value of $u_-<u<u_+$, there are two solutions for $\cos\theta$ with the opposite signs. According to equation \eqref{eq:psipl02}, $\cos\theta>0$ corresponds to $u'>0$, and $\cos\theta<0$ to $u'<0$. Accordingly, there are two solutions $\theta_{\pm}$ for the angle $\theta$,
\beq
\theta_{\pm}(\tau)=\text{arccos}\,\left(\pm\sqrt{1-\frac{C(u^2(\tau)+1)}{2u(\tau)}}
\right)
\label{theta_sol}
\eeq
The trajectories $\theta_{\pm}(\psi)$ corresponding to these two solutions are shown in figure \ref{fig:phase_portrait_S4}. For $u=u_{\pm}$, $\theta=\pi/2$. 
Functions $\psi(\tau)$ and $\theta(\tau)$
are illustrated in figure~\ref{fig:angles}a,b for $C=0.10$ and $C=0.74$.
 
\subsection{Solution for $\phi(\tau)$}

The solution for $\phi(\tau)$ follows from integrating equation \eqref{eq:phipl02}, with the integrand expressed in terms of the function of time, $u(\tau)=u_-+\delta\,\mathrm{cn}^2(\zeta,k)$, determined in equation \eqref{sol_eq}.
The full derivation, based on a partial-fraction decomposition of the integrand, is given in
Appendix~\ref{app:phi_derivation}. Here we will present the result, using the following definitions of the special
functions entering the result, together with the explicit value of the elliptic
characteristic.

The Jacobi amplitude $\mathrm{am}(\zeta,k^2)$ is defined by
\begin{equation}
    \mathrm{am}(u, k^2)=\varphi \quad\Longleftrightarrow\quad u=F(\varphi, k^2)=\int_0^\varphi\frac{\mathrm{d}\vartheta}{\sqrt{1-k^2\sin^2\vartheta}},
    \label{am_def}
\end{equation}
where $F(\varphi, k^2)$ is the incomplete elliptic integral of the first kind
(\cite{Byrd1954Handbook, Chandrasekharan1985}), and
\begin{equation}
    \Pi(n; \varphi| k^2)=\int_0^\varphi\frac{\mathrm{d}\vartheta}{(1-n\sin^2\vartheta)\sqrt{1-k^2\sin^2\vartheta}}
    \label{eq:Pi}
\end{equation}
is the incomplete elliptic integral of the third kind (\cite{Byrd1954Handbook}).

In terms of the turning points $u_\pm$ (Eq.~\eqref{pm}) and $\delta=u_+-u_-$
(Eq.~\eqref{defi}), the elliptic characteristic entering the solution is given
explicitly by
\begin{equation}
   n_{+}=\delta\,\frac{u_+ + i}{1+u_+^2},
   \label{eq:el_char}
\end{equation}
which is genuinely complex ($\mathrm{Im}\,n_{+}\neq 0$) whenever $0<C<1$, since then
$u_+\in\mathbb R$ and $\delta\neq 0$. With the above definitions,  the closed-form
solution reads
\begin{equation}
    \phi(\tau)=\phi_0-\frac{2}{\Omega}\mathrm{Re}\left\{\frac{n_+}{\delta}\left[\Pi(n_+; \mathrm{am}(\zeta,k^2)| k^2)-\Pi(n_+; \mathrm{am}(\zeta_0,k^2)| k^2)\right] \right\},
    \label{eq:phi_sol}
\end{equation}
where $\zeta=\Omega(\tau-\tau_0)$, as in Eq.~\eqref{defi}. As shown in
Appendix~\ref{app:phi_derivation}, the quantity $n_+$ in \eqref{eq:el_char} arises
together with its complex conjugate $n_-=n_+^{*}$ from a partial-fraction
decomposition of the integrand of \eqref{eq:phipl02}; the two conjugate
contributions combine into the real quantity written as $\mathrm{Re}\{\cdots\}$ in
\eqref{eq:phi_sol}.

Since $n_+$ is complex, the incomplete elliptic integral of the third kind
$\Pi(n_{+};\varphi|k^2)$ appearing in \eqref{eq:phi_sol} must be evaluated for a
complex parameter. The requisite analytic continuation in $n$ is well defined
provided $1 - n\sin^2\vartheta$ has no zero on the real integration path; a
rigorous treatment of the branch structure for complex characteristics is given by
\citet{ArmitageEberlein2006}. While the Legendre form \eqref{eq:Pi} provides a
convenient analytic definition by continuation in $n$, a robust numerical evaluation
is obtained by rewriting $\Pi$ in Carlson's symmetric forms
\citep{Carlson1977Special, Carlson1995Complex, CarlsonDLMF2020EllipticIntegrals}.
This avoids explicit branch tracking in the integrand and is well-suited for complex parameters.

In our implementation we evaluate $\Pi(n; \varphi|\, k^2)$ for $\varphi\in[0,\pi/2]$,
and extend it to $\varphi\in(\pi/2,\pi]$ using the identity
\begin{equation}
\Pi(n;\varphi|k^2)=2\Pi(n|k^2)-\Pi(n;\pi-\varphi|k^2),
\qquad \varphi\in\Bigl(\frac{\pi}{2},\pi\Bigr],
\label{eq:Pi_reflection}
\end{equation}
with $\Pi(n|k^2)\equiv\Pi(n;\pi/2|k^2)$. The azimuth $\phi(\tau)$ is
then computed from the closed form \eqref{eq:phi_sol} with $\varphi=\mathrm{am}(\zeta|k^2)$
obtained from the Jacobi functions. Unlike a direct numerical quadrature of
\eqref{eq:phipl02}, the closed form \eqref{eq:phi_sol} evaluates $\phi(\tau)$ through
well-characterised special functions, so its accuracy can be systematically controlled
(via established implementations and identities such as the connection formula), while
preserving the exact periodic-drift structure of the solution, given in
equation~\eqref{phiphiperp}.

\subsection{Connection formula and the drift of $\phi(\tau)$}
\label{sec:drift}

The angles $\psi(\tau)$ and $\theta(\tau)$ are periodic with the period
$T=2K(k^2)/\Omega$. Consequently, $\phi'(\tau)=-\sin 2\psi(\tau)=-2u/(1+u^2)$ is also $T$-periodic. In general, however, $\phi(\tau)$ itself is not periodic: it acquires a constant increment after each period $T$.

To compute this increment from \eqref{eq:phi_sol} one must specify the analytic
continuation of the incomplete elliptic integral of the third kind. For real $z$ one has
\[
\mathrm{am}(\zeta+2K(k^2),k^2)=\mathrm{am}(\zeta,k^2)+\pi,
\]
which is consistent with $\mathrm{cn}(\zeta+2K(k^2),k)=-\mathrm{cn}(\zeta,k)$ and $\mathrm{cn}(\zeta,k)=\cos(\mathrm{am}(\zeta,k^2))$.
We then use the connection formula (a consequence of the
$\pi$-periodicity of the integrand in the definition of $\Pi$):
\begin{equation}
\Pi\!\big(n;\,\varphi+\pi\,\big|\,k^2\big)=\Pi\!\big(n;\,\varphi\,\big|\,k^2\big)+2\,\Pi(n|k^2),
\qquad
\Pi(n|k^2)\equiv \Pi\!\big(n;\tfrac{\pi}{2}\big|k^2\big).
\label{eq:legendre_connection_Pi}
\end{equation}
Although \eqref{eq:legendre_connection_Pi} is often stated for real $n$, it remains valid for complex $n$ by analytic continuation: for fixed $k^{2}\in(0,1)$ the function $\Pi(n;\varphi|k^{2})$ is analytic in $n$ as long as the denominator $1-n\sin^{2}\vartheta$ has no zero on the real integration path in \eqref{eq:Pi}. 
For $\varphi=\pi/2$ this obstruction occurs precisely when $n$ lies on the real interval $[1,\infty)$, where a pole crosses the contour; this interval is the \emph{standard branch cut} in the $n$-plane used to define the principal branch of $\Pi$ (the value on one side of the cut differs from the value on the other by a discontinuity). 
Since in our case $\mathrm{Im}\,n_{+}\neq 0$, the characteristic stays away from this cut and the continuation is unambiguous, so the identity \eqref{eq:legendre_connection_Pi} applies without modification.
Applying \eqref{eq:legendre_connection_Pi} to \eqref{eq:phi_sol} yields a constant
phase increment of $\phi$ after one period $T$:
\begin{eqnarray}
\Delta\phi \equiv \phi(\tau+T)-\phi(\tau)
&=& -\frac{2}{\Omega}\mathrm{Re}\!\left\{\frac{n_+}{\delta}\Big[\Pi\!\big(n_+;\mathrm{am}(\zeta+2K, k^2)|k^2\big)-\Pi\!\big(n_+;\mathrm{am}(\zeta, k^2)|k^2\big)\Big]\right\}\nonumber\\
&=& -\frac{4}{\Omega}\mathrm{Re}\!\left\{\frac{n_+}{\delta}\,\Pi(n_+|k^2)\right\},
\label{eq:delta_phi}
\end{eqnarray}
independent of $\tau$ ($n_+$ is given in Eq. \eqref{eq:el_char} and $\zeta$ is related to $\tau$ by Eq. \eqref{defi}).

Therefore, $\phi(\tau)$ can be written as a sum of a linear drift and a periodic part, $\phi(\tau)=\Delta \phi\, \tau/T+\phi_{\rm per}(\tau),$ with $
\phi_{\rm per}(\tau+T/2)=\phi_{\rm per}(\tau)$, as already pointed out in Eq.~\eqref{phiphiperp}. The time scale $\tau_{\rm drift}$ of the rotation around $Z$ is related to the mean drift rate of $\phi$,
\begin{equation}
\tau_{\rm drift}=\frac{2\pi T}{|\Delta \phi|}=\frac{\pi K(k^2)}{|\mathrm{Re}\!\left\{\tfrac{n_+}{\delta}\,\Pi(n_+|k^2)\right\}|}.
\label{eq:tau_drift}
\end{equation}
The plot of $\tau_{\rm drift}$ as a function of $C$ is shown in the figure~\ref{fig:TvsC}.

Finally, the orientation becomes strictly periodic only if the drift is
commensurate with $2\pi$, i.e. if $\Delta\phi/(2\pi)\in\mathbb{Q}$.
If $\Delta\phi/(2\pi)=p/q$ (in lowest terms), then the full Euler-angle triplet
repeats after $T_{\rm orient}=q\,T$. Otherwise, the motion is quasiperiodic. Equivalently, the dynamics may be viewed as a linear flow on a two-torus \( S^{1} \times S^{1} \), where one angular coordinate is the phase of the \( T \)-periodic \( (\theta,\psi) \)-motion and the other is the azimuthal angle \( \phi \) taken modulo \( 2\pi \). In this representation, rational \( \Delta\phi/(2\pi) \) yields a closed orbit on the torus, whereas irrational
\( \Delta\phi/(2\pi) \) gives a non-closing orbit that is dense on it \citep{Arnold1988}.

Exact solutions for the Euler angles $\psi(\tau)$, $\theta(\tau)$, $\phi(\tau)$ mod $2\pi$, and $\phi_{\rm per}(\tau)$ for $C\!=\!0.10$ and $C\!=\!0.74$ are plotted in figure \ref{fig:angles}. These angles are also evaluated numerically by integrating equations \eqref{eq:thetapl02}-\eqref{eq:phipl02} with a high-accuracy  \textsc{SciPy}'s  implementation of the
DOP853 algorithm (i.e. explicit Runge-Kutta method of order 8) from \cite{Hairer2008}. The
maximum absolute deviation between the exact solutions (solid lines) and numerical solutions (symbols) over the plotted interval is of the order of $10^{-11}$.

In the regime $0<C<1$, the period $T(C)$ of our analytic solution (the common period of $\psi(\tau)$ and $\theta(\tau)$) exhibits the same limiting behavior as the period $T_y$ reported by \cite{joshi} for $C_{2v}$ symmetry, with two in general different mobility coefficients $\mu_a$ and $\mu_b$ (their Figs.~4-5). Their period $T_y$ remains finite for $0<C<1$, but diverges upon approaching the separatrix. In our parametrization, and for $\mu_a=\mu_b$, the period
$T(C)=2K(k^2)/\Omega$ tends to $\sqrt{2}\,\pi$ as $C\to 1^{-}$ and grows without bound as $C\to 0^{+}$, in agreement with their finite small-oscillation period
$T_{y m}\sqrt{\mu_a\mu_b}=\sqrt{2}\pi$ near the centers and the logarithmic divergence of $T_y$ near the heteroclinic connection. The same qualitative dependence of the period on the
amplitude of the orientational oscillation was reported by
\citet{vaquero} for U-shaped disks (their
Fig.~5\textit{a}), where the period was extracted from
direct numerical integration of the resistance-matrix
equations. Our closed-form expression~\eqref{period}
for the special case with $\mu_a=\mu_b$ provides the exact functional dependence that their
numerical data approximate. Moreover, the drift time
$\tau_{\rm drift}=\pi K(k^2)/|\mathrm{Re}\{\tfrac{n_+}{\delta}\,\Pi(n_+|k^2)\}|$  provides a natural analogue of their Floquet rotation measure: the dimensionless drift accumulated over one oscillation,
$\Delta\phi/2\pi=T/\tau_{\rm drift}$ (equivalently, the complementary quantity $1-T/\tau_{\rm drift}$) is bounded in the same way as the frequency ratio in their Fig.~5, namely
\[
1-\frac{1}{\sqrt{2}}\ \le\ 1-\frac{T}{\tau_{\rm drift}}\ \le\ \frac{1}{2}
\qquad\text{(i.e.\ }0.2929\ldots \le 1-\tfrac{T}{\tau_{\rm drift}}\le 0.5\text{)},
\]
or, equivalently,
\beq
\frac{1}{2}\ \le\ \frac{T}{\tau_{\rm drift}}\ \le\ \frac{1}{\sqrt{2}}.\label{ratio}
\eeq
This links our exact drift-period decomposition to their frequency-ratio characterization of the Poincar\'e map. 

We note that \citet{joshi} characterised the drift-period
relationship through a Poincar\'e map, extracting the
frequency ratio numerically for specific shapes computed by
the boundary integral method. For $\mu_a=\mu_b$, the closed-form expressions
\eqref{eq:delta_phi} and~\eqref{eq:tau_drift} derived here
replace this numerical extraction with explicit functions of
$C$, enabling the frequency ratio to be evaluated to
arbitrary precision for any value of the first integral
without time integration.

\section{Motion of the particle's centre of mass}
\label{sec:cent_mass}
\subsection{Special cases}\label{sec:cm_special}
 
For $C=1$, the  centre of mass moves vertically with a constant speed (see equations~\ref{eq:}, \ref{C=1}), 
\beq
X_{\rm cm}(\tau)=X_{\rm cm}(0),\hspace{1cm} Y_{\rm cm}(\tau)=Y_{\rm cm}(0), \hspace{1cm} Z_{\rm cm}(\tau)=-\frac{\mu_3}{\mu_b}\tau+Z_{\rm cm}(0).\eeq

For $C=0$, the special solutions \eqref{theta_exp}-\eqref{phi_const} for the Euler angles give $\psi(\tau)=k\tfrac{\pi}{2}$ ($k\in\{0,1,\dots, 4\}$); the azimuthal angle is locked, $\phi(\tau)\equiv\phi_0$, while $\tan(\theta(\tau)/2)=A\exp(-s\tau)$, with  $s:=(-1)^k$ and $A=\tan(\theta_0/2)$.
Substituting these elementary Euler-angle solutions into the translational kinematics yields explicit centre-of-mass trajectories,
\begin{align}
X_{\rm cm}(\tau)
&=X_{\rm cm}(0)+\frac{4\mathcal K}{s}
\left(\frac{A e^{-s\tau}}{1+A^2 e^{-2s\tau}}-\frac{A}{1+A^2}\right)\sin\phi_0,
\\[2pt]
Y_{\rm cm}(\tau)
&=Y_{\rm cm}(0)-\frac{4\mathcal K}{s}
\left(\frac{A e^{-s\tau}}{1+A^2 e^{-2s\tau}}-\frac{A}{1+A^2}\right)\cos\phi_0,
\\[2pt]
Z_{\rm cm}(\tau)
&=Z_{\rm cm}(0)-\frac{\mu_3}{\mu_b}\,\tau
+\frac{4\mathcal K}{s}\left(\frac{1}{1+A^2 e^{-2s\tau}}-\frac{1}{1+A^2}\right).
\end{align}
Thus, the horizontal motion is purely one-dimensional: $(X_{\rm cm},Y_{\rm cm})$ evolves along a straight line whose direction is fixed by $\phi_0$, and the horizontal displacement saturates as $\sin\theta(\tau)\to 0$. Vertically, $Z_{\rm cm}$ consists of the uniform drift $-\mu_3/\mu_b$ plus a transient correction that decays as $\theta(\tau)$ approaches its limiting value. In 3D, the centre-of-mass moves in the single vertical plane.

\subsection{Vertical velocity $Z'_{\rm cm}(\tau)$ and vertical displacement $Z_{\rm cm}(\tau)$}
\label{sec:z_vel}

The vertical component of the velocity $Z'_{\rm cm}$ is independent
of the azimuth $\phi$ and reads 
\beq Z'_{\rm cm}=\mathcal{K}(-\mathcal{V}_0+2 \sin^2 \theta),\label{Zprime}\eeq
as shown in Eq.~\eqref{eq:}. 
In the regime $0 < C < 1$, the orientation dynamics admits
closed orbits in the $(\psi, \theta)$ phase plane, shown in figure \ref{fig:phase_portrait_S4}, and the
angles $\psi(\tau)$ and $\theta(\tau)$ are $T$-periodic,
with
$
T={2K(k^2)}/{\Omega},$ 
as shown in Eq.~\eqref{period}. Therefore, from equation \eqref{Zprime} it follows that the vertical centre-of-mass velocity \(Z'_{\rm cm}(\tau)\), similarly as  $\sin^2\!\theta(\tau),$ is periodic with the period $T/2$. 
Since on a closed orbit with \(0<C<1\) one has
\(\sin^2\theta(\tau)\in[C,1]\), the instantaneous velocity $Z'_{\rm cm}(\tau)$ satisfies the bounds
\begin{equation}
\mathcal K\bigl(-\mathcal V_0+2C\bigr)
\le
Z'_{\rm cm}(\tau)
\le
\mathcal K\bigl(-\mathcal V_0+2\bigr),
\qquad \tau\in[0,T].
\label{eq:VZ_inst_bounds_simple_new}
\end{equation}

\begin{figure}
    \centering
    \includegraphics[width=0.95\linewidth]{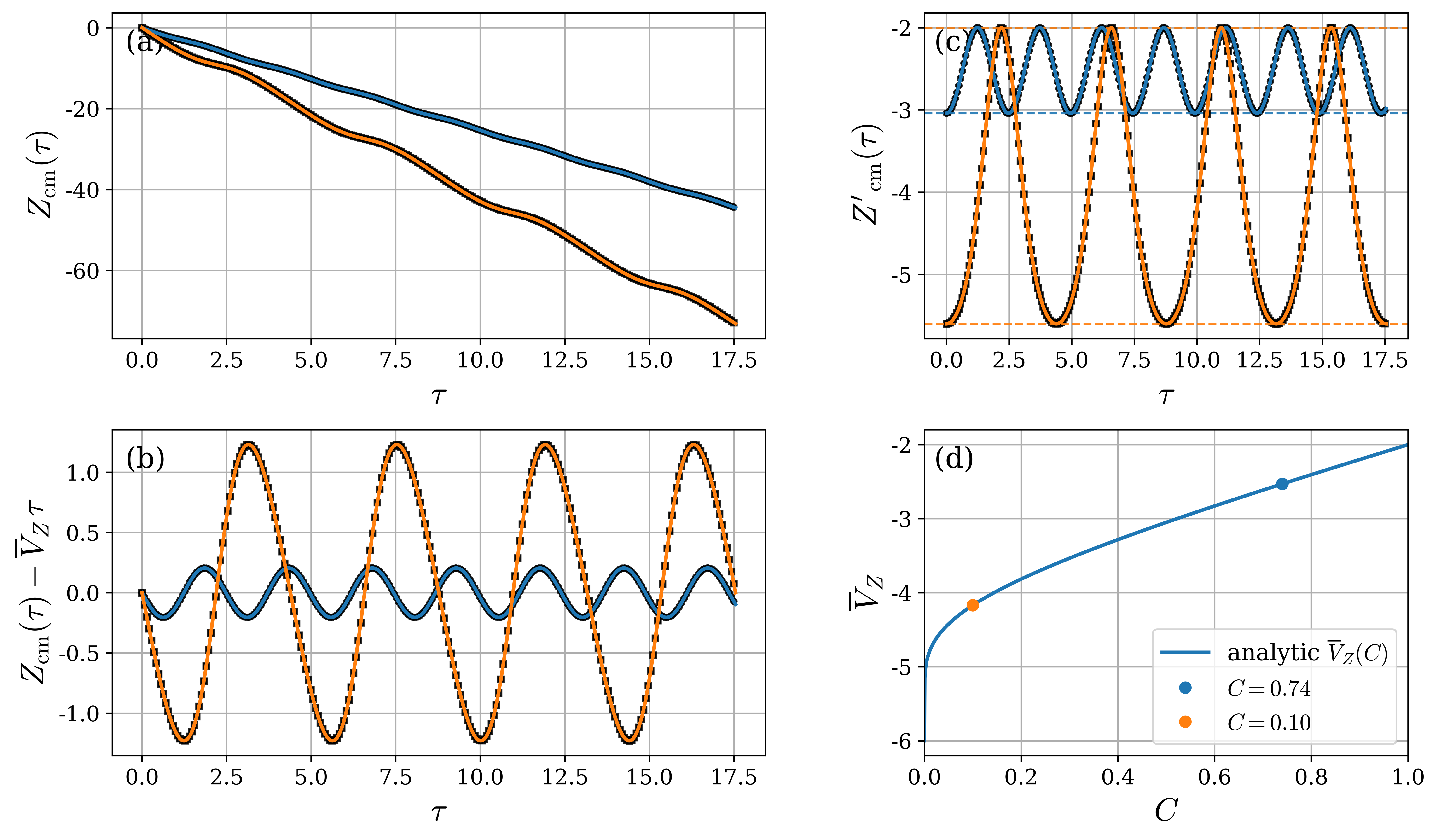}
    \vspace{-0.3cm}
    \caption{
Exact solution for vertical projection of the centre-of-mass dynamics at $\mu_1\!=\!1.0$, $\mu_3\!=\!3.0$, and $\mu_b\!=\!0.5$ (solid colour curves).
Panels (a), (b) and (c), respectively: $Z_{\rm cm}(\tau)$; $T/2$-periodic $Z_{\rm cm}(\tau)\!-\!\overline{V}_Z\,\tau$; and $T/2$-periodic velocity $Z'_{\rm cm}(\tau)$ for two representative closed orbits, $C\!=\!0.74$ (blue) and $C\!=\!0.10$ (orange).
High-accuracy numerical integration using SciPy's DOP853 solver is also shown (black markers).
Panel (d): 
The exact orbit-averaged sedimentation velocity $ -\mu_3/\mu_b\!<\!\overline{V}_Z(C) \!<\!-\mu_1/\mu_b$ as a function of the first integral $C$.} 
    \label{fig:cm_motion}
\end{figure}

Using the first integral $C=\sin^2\theta\sin2\psi$ and the
substitution $u=\tan \psi$, we obtain $\sin^2\theta=\tfrac{C}{2}\frac{1+u^2}{u}$, so that
\begin{equation}
    Z'_{\rm cm}= \mathcal{K}\left[-\mathcal{V}_0+C \left(u(\tau) +u^{-1}(\tau)\right) \right],
    \label{eq:z_vel}
\end{equation}
where $u(\tau)$ is given by the Jacobi solution in Eq.~\eqref{sol_eq}. 
The time-dependent $ Z'_{\rm cm}$ is shown in Fig.~\ref{fig:cm_motion}c.

We obtain $Z_{\rm cm}(\tau)$ by integrating Eq.~\eqref{eq:z_vel} over time. 
The integral contains standard quadratures of Jacobi
elliptic functions. The details of this reduction 
are given in Appendix~\ref{app:Zcm_derivation}. The resulting
integrals are expressed via the incomplete elliptic integral of the second kind,
\begin{equation}
    E(\varphi\, |\, k^2)=\int_{0}^\varphi \sqrt{1-k^2\sin^2\vartheta}\,\mathrm{d}\vartheta,
    \label{eq:E_def}
\end{equation}
and via the incomplete elliptic integral of the third kind, $\Pi(n;\varphi\,|\,k^2)$
(Eq.~\eqref{eq:Pi}), evaluated at the real characteristic $n=k^2$. With
$\mathrm{am}(\zeta,k^2)$ as defined in Eq.~\eqref{am_def}, and $\zeta=\Omega(\tau-\tau_0)$
as in Eq.~\eqref{defi}, the closed-form vertical displacement reads
\begin{eqnarray}
 Z_{\rm cm} (\tau) =Z_{\rm cm}(0) +\frac{\mathcal{K}}{\Omega}\Bigg[&-&\mathcal{V}_0(\zeta-\zeta_0)+Cu_+\left(E(\mathrm{am}(\zeta, k^2)|k^2)-E(\mathrm{am}(\zeta_0, k^2)|k^2)\right)\nonumber\\
 &+&\frac{C}{u_+}\left(\Pi(k^2;\,\mathrm{am}(\zeta, k^2) \,|\, k^2)-\Pi(k^2;\,\mathrm{am}(\zeta_0, k^2) \,|\, k^2)\right)\bigg].
 \label{Zcm}
\end{eqnarray}
The resulting closed-form expression for $Z_{\rm cm}(\tau)$ is illustrated in
Fig.~\ref{fig:cm_motion}a, while Fig.~\ref{fig:cm_motion}b shows that subtracting the mean
drift, $Z_{\rm cm}(\tau)-\overline{V}_Z\,\tau$, leaves a purely periodic vertical
displacement with the period $T/2$. The average sedimentation velocity $\overline{V}_Z$ will be evaluated exactly in the next section.

\subsection{Average sedimentation velocity}\label{sec:av_sed_vel}
The practice of averaging instantaneous sedimentation quantities over orientational cycles has a long history in low-Reynolds-number hydrodynamics
\citep{Lamb1932,HappelBrenner1983}. Since the instantaneous sedimentation velocity $Z'_{\rm cm}(\tau)$  oscillates along the orbit, it is natural to
characterize the long-time vertical drift by its average over one orientation period. This yields a quantity depending only on the first integral \(C\).

At this stage, to evaluate the mean vertical sedimentation velocity $\overline{V}_Z$, it is most convenient to use the closed-form expression for \(Z_{\rm cm}(\tau)\)
derived in Appendix~\ref{app:Zcm_derivation} (Eq.~\eqref{Zcm}), rather than to
re-express the average through a separate integration of \(\sin^2\theta(\tau)\). Therefore, 
\begin{equation}
\overline{V}_Z
=
\frac{1}{T/2}\int_0^{T/2} Z'_{\rm cm}(\tau)\,{\rm d}\tau
=
\frac{Z_{\rm cm}(\tau+T/2)-Z_{\rm cm}(\tau)}{T/2}.
\label{eq:Vz_mean_def_new}
\end{equation}
The
reduction of Eq.~\eqref{eq:Vz_mean_def_new} to a closed algebraic expression in $C$,
via the periodicity increments of the Jacobi amplitude $\mathrm{am}(\zeta,k^2)$ under
$\zeta\mapsto\zeta+2K(k^2)$, and the identity relating the complete integrals
$\Pi(k^2|k^2)$ and $E(k^2)$, is given in full in Appendix~\ref{app:Vz_derivation}.
The result is
\begin{equation}
\overline{V}_Z
=
\mathcal K\left[
-\mathcal V_0
+
2\left(1+\sqrt{1-C^2}\right)\frac{E(k^2)}{K(k^2)}
\right],
\label{eq:mean_vel}
\end{equation}
where $E(k^2)\equiv E(\pi/2|k^2)$ (defined in equation \eqref{eq:E_def}), $K(k^2)$ (defined in equation \eqref{defK}) denote the complete
elliptic integrals of the second and first kind, respectively, and we remind that

\[
k^2=\frac{2\sqrt{1-C^2}}{1+\sqrt{1-C^2}},
\qquad
\mathcal K=\frac{\mu_3-\mu_1}{2\mu_b},
\qquad
\mathcal V_0=\frac{2\mu_3}{\mu_3-\mu_1}.
\]

The behaviour of \(\overline{V}_Z(C)\) is shown in Fig.~\ref{fig:cm_motion}d for exemplary values of $\mu_1$, $\mu_3$, $\mu_b$ and $C$.
In the limit \(C\to 1^{-}\), one has \(k^2\to 0\) and therefore \(E(k^2)/K(k^2)\to 1\). It follows
that
\[
\overline{V}_Z \to \mathcal K(-\mathcal V_0+2)=-\frac{\mu_1}{\mu_b}\hspace{1cm}\mbox{for $C\to 1^{-}$},
\]
which agrees with the centre solution \(\theta=\pi/2\). In the opposite limit \(C\to 0^{+}\),
one has \(k^2\to 1\), while \(E(k^2)/K(k^2)\to 0\), hence
\[
\overline{V}_Z\to -\mathcal K\mathcal V_0=-\frac{\mu_3}{\mu_b}\hspace{1cm}\mbox{for $C\to 0^{+}$}.
\]
Therefore, the orbit-averaged sedimentation velocity varies between the two limiting values
\(-\mu_3/\mu_b\) and \(-\mu_1/\mu_b\), consistently with the behaviour shown in
Fig.~\ref{fig:cm_motion}d. Note that in our normalisation, according to equation \eqref{tauo}, velocity unit is equal to $\mu_b |\boldsymbol{F}|/(\pi\eta L)$. Therefore,  the dimensional velocity does not depend on $\mu_b$; it depends only on the translational-translational mobility coefficients, $\mu_1$ and $\mu_3$. It is also useful to discuss the limiting values of the instantaneous vertical velocity $Z'_{\rm cm}$, plotted in figure \ref{fig:cm_motion}c. In the whole range of $0 < C < 1$, the maximum value of $Z'_{\rm cm}$, attained at the horizontal position with $\theta=\pi/2$, is equal to $-\mu_1/\mu_b$. However, the minimal value is larger than $-\mu_3/\mu_b$, because the particle, while moving along its trajectory, does not reach a vertical position, having $0 < \theta < \pi$.

\subsection{XY-plane projection of the centre-of-mass motion}
\label{sec:xy_plane}

The solution of the orientation dynamics is available in closed form (see Eqs. \eqref{sol_eq}, \eqref{theta_u} and \eqref{eq:phi_sol}). In the regime $0<C<1$, the angles $\psi(\tau)$ and $\theta(\tau)$ are strictly $T$-periodic, whereas the azimuth $\phi(\tau)$ generally contains a secular drift plus a $T/2$-periodic modulation. The horizontal centre-of-mass motion inherits this mixed periodic/quasiperiodic structure and generically forms rosette-like trajectories in the $XY$ plane, confined between two co-centred circles ("envelopes") of radii $\rho_{\rm in} < \rho_{\rm out}$, as illustrated in figure \ref{fig:xy_pane_mot}. In this section, we develop a  description of these rosettes, determine their inner and outer envelope radii, $\rho_{\rm in}$ and $\rho_{\rm out}$, and explain the origin of the sharp tips (cusps), visible in figure \ref{fig:xy_pane_mot}. We show that the horizontal motion, like the vertical one, admits an exact closed-form solution for $X_{\rm cm}$ and $Y_{\rm cm}$, which allows for analytical description of the trajectory.

\subsubsection*{Kinematics}
The planar components of \eqref{eq:} in plane $XY$ are
\begin{equation}
(X'_{\rm cm}(\tau), Y'_{\rm cm}(\tau))
=\mathcal{K}\,\sin 2\theta(\tau)\,(-\sin\phi(\tau), \cos \phi(\tau)).
\label{eq:XY_components}
\end{equation}

Substituting the analytic expressions for $\theta(\tau)$ and $\phi(\tau)$ into
\eqref{eq:XY_components} leads to quadratures of the type
\begin{equation}
\begin{pmatrix}
    X_{\rm cm}(\tau)-X_{\rm cm}(0)\\
    Y_{\rm cm}(\tau)-Y_{\rm cm}(0)
\end{pmatrix}=
\mathcal{K}\int_0^\tau \sin 2\theta(\tau')\,
\begin{pmatrix}
    -\sin\phi(\tau')\\
    \cos\phi(\tau')
\end{pmatrix}\,{\rm d}\tau',
\label{eq:XY_quadratures}
\end{equation}
whose integrands are products of a Jacobi-elliptic function of $\tau$ (via
$\theta(\tau)$) and trigonometric functions of $\phi(\tau)$, the latter expressed
through an incomplete elliptic integral of the third kind with a complex
characteristic. Although these quadratures do not belong to the standard classes of
elliptic integrals, they will be shown below to admit an exact closed form,
Eq.~\eqref{eq:Xic_closed_form}.

\begin{figure}
    \centering
    \includegraphics[width=0.9\linewidth]{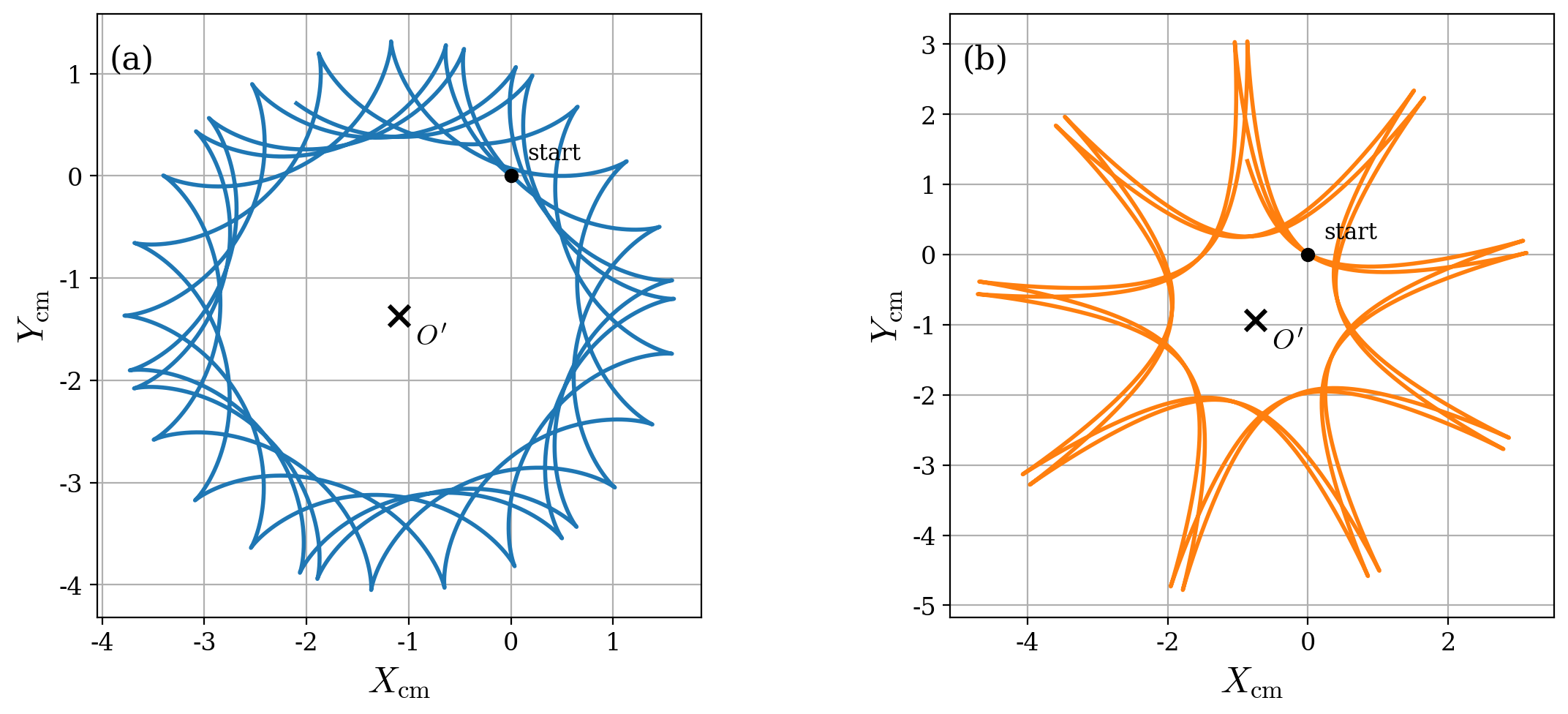}\vspace{-0.3cm}
    \caption{Projection of the centre-of-mass trajectories onto the horizontal plane, obtained by direct numerical integration (SciPy DOP853) of Eq.~\eqref{eq:XY_quadratures}, using the closed forms of $\theta(\tau)$ and $\phi(\tau)$ 
    for $\mu_1=1.0$, $\mu_3=3.0$, and $\mu_b=0.5$. Panels (a) and (b) show the projected paths $(X_{\rm cm}(\tau),Y_{\rm cm}(\tau))$ for two representative orbits with first integral $C=0.74$ and $C=0.10$, respectively. The black dot marks the initial position. The initial values of the angles are \(\phi_0=0.9\), \(\psi_0=\pi/4\), and \(\theta_0=\sin^{-1}\sqrt{C}<\pi/2\).
    }
    \label{fig:xy_pane_mot}
\end{figure}

It is convenient to separate the \emph{instantaneous speed} from the \emph{instantaneous velocity direction}.
From \eqref{eq:XY_components} we obtain
\begin{equation}
V_{\rm hor}(\tau):=\sqrt{(X'_{\rm cm})^{2}+(Y'_{\rm cm})^{2}}
=|\mathcal{K}\,\sin 2\theta(\tau)|,
\label{eq:Vhor_again}
\end{equation}
and the ``heading direction'' is defined by the unit vector parallel to velocity in the $XY$ plane,
\begin{equation}
\hat{\bm t}(\tau)=\frac{\left(X'_{\rm cm}(\tau), Y'_{\rm cm}(\tau)\right)}{V_{\rm hor}}
=\mathrm{sgn}(\mathcal{K}\sin 2\theta(\tau))\,(-\sin\phi(\tau),\ \cos\phi(\tau)).
\label{eq:heading}
\end{equation}
Thus, for a given particle shape, the speed of the horizontal motion is set solely by $\sin 2\theta(\tau)$,
which is $T/2$-periodic (for $0<C<1$), while the direction rotates and oscillates 
with $\phi(\tau)$.
This is the basic mechanism behind rosette formation:  the motion along the trajectory alternates between faster and slower velocity values in a strictly periodic way, while the direction of the planar displacement slowly precesses while oscillating. As a result, each half-period $T/2$ draws a similar ``petal-like motif'', that is later repeated at a rotated position owing to the accumulated azimuthal drift. 
Typical projections of the centre-of-mass trajectories
on the $XY$ plane are shown in figure~\ref{fig:xy_pane_mot} for $C=0.1$ and $C=0.74$. The curves were obtained by integrating \eqref{eq:XY_components} with SciPy's DOP853 solver, while evaluating $\theta(\tau)$, $\psi(\tau)$ and $\phi(\tau)$ from the closed-form expressions
derived in Sec.~\ref{sec:exact_euler}. 

Figure \ref{fig:xy_pane_mot} illustrates formation of a rosette around a certain point $O'$ that is shifted with respect to the point $O$ corresponding to the initial horizontal position $\left(X_{\rm cm}(0), Y_{\rm cm}(0)\right)=\bm{0}$ of the particle centre of mass. It is clearly visible that oscillations of the centre of mass position are superimposed with the rotation around $O'$. However, even a long-time numerical integration of equation \eqref{eq:XY_components}  is not sufficient to determine with high precision
the position of $O'$ (and therefore also the envelope radii $\rho_{\rm in}$ and $\rho_{\rm out}$). Therefore, this method, based on averaging, is not efficient, especially for small values of $C$, when both $T$ and $\tau_{\rm drift}$ are very large. This motivates the alternative strategy adopted below instead of evaluating numerically the integral \eqref{eq:XY_quadratures} to determine $X_{\rm cm}(\tau)$ and $Y_{\rm cm}(\tau)$.  

We describe the planar motion using two characteristic radii, the inner and outer envelope radii $\rho_{\rm in}$ and $\rho_{\rm out}$, which do not depend on how the trajectory is centred in the $XY$ plane. 
Both are obtained below in closed form. It is also useful to recall the constant angular velocity of the azimuthal drift,
\begin{equation}
\omega_Z=\frac{\Delta\phi}{T}, \quad \mbox{and}\quad |\omega_Z|=\frac{2\pi}{\tau_{\rm drift}},
\end{equation}
which defines the frame co-rotating with the constant angular velocity, used below to exhibit the periodic structure of the orbit.
Here \(T\) is the orientation period of \(\psi(\tau)\) and \(\theta(\tau)\), while
\(\Delta\phi\) is the net azimuthal increment accumulated over one period.

\subsubsection*{Closed-form solution and the envelope radii $\rho_{\rm in/out}$}

As shown in Eq.~\eqref{phiphiperp}, and later also in Sec.~\ref{sec:exact_euler}, the azimuthal angle can be decomposed into
a uniform drift and a periodic part,
\beq
\phi(\tau)=\omega_Z \tau+\phi_{\rm per}(\tau), \;\; \mbox{ with }\;\;
\phi_{\rm per}(\tau+T/2)=\phi_{\rm per}(\tau).
\label{eq:phi_decomp_fourier}
\eeq

To separate the uniform rotation from the intrinsic periodic modulation of the planar motion,
it is convenient to combine the horizontal coordinates into the complex variable
\begin{equation}
\Xi(\tau):=X_{\rm cm}(\tau)+{\rm i}Y_{\rm cm}(\tau).
\label{eq:Xi_def_new}
\end{equation}
Equation \eqref{eq:XY_components} then takes the form
\begin{equation}
\Xi'(\tau)= {\rm i}\,\mathcal{K}\,\sin 2\theta(\tau)\,e^{{\rm i}\phi(\tau)}.
\label{eq:Xi_prime_new}
\end{equation}

\begin{figure}
    \centering
    \includegraphics[width=0.95\linewidth]{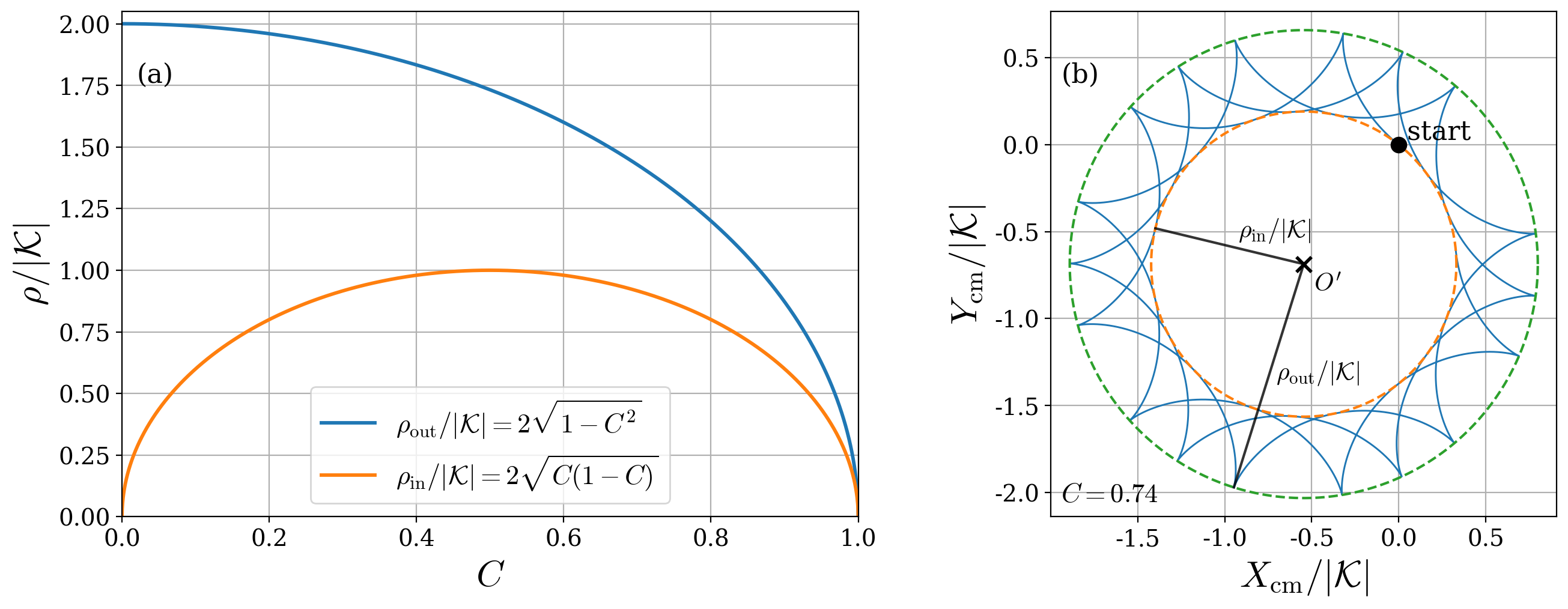}
    \vspace{-0.4cm}
    \caption{Inner and outer radial envelopes ($\rho_{\rm in}$ and $\rho_{\rm out}$) of the horizontal component of the centre-of-mass motion.
(a) Normalized radii $\rho_{\rm in}/|\mathcal{K}|$ and $\rho_{\rm out}/|\mathcal{K}|$ as functions of the first integral $C$.
(b) Laboratory-frame trajectory $(X_{\rm cm}(\tau),Y_{\rm cm}(\tau))$ over 9 orientation periods $T$, for $C=0.74$. Dashed circles indicate the radii $\rho_{\rm in}/|\mathcal{K}|$ and $\rho_{\rm out}/|\mathcal{K}|$. The initial values of the angles are \(\phi_0=0.9\), \(\psi_0=\pi/4\), and \(\theta_0=\sin^{-1}\sqrt{C}< \pi/2\).}
    \label{fig:rho_in_out_vs_C}
\end{figure}

Equation \eqref{eq:Xi_prime_new} can be integrated exactly. We will show that a particular solution of Eq. \eqref{eq:Xi_prime_new} is
\begin{equation}
\Xi_{\rm c}(\tau)=\mathcal K\,A(\tau)\,e^{{\rm i}\phi(\tau)},
\qquad
A(\tau):=\sin 2\theta\,\sin 2\psi-2{\rm i}\,\sin\theta\,\cos 2\psi .
\label{eq:Xic_closed_form}
\end{equation}
Differentiating and using $\phi'=-\sin2\psi$ gives
$\Xi_{\rm c}'=\mathcal K\bigl(A'-{\rm i}\sin2\psi\,A\bigr)e^{{\rm i}\phi}$, where the
derivative of the phase factor $\phi$ supplies the second term. With
$\theta'=-\cos2\psi\sin\theta$ and $\psi'=\sin2\psi\cos\theta$ one obtains
\begin{align*}
A' &= 2\sin\theta\sin 2\psi\cos 2\psi
+{\rm i}\,\sin 2\theta\left(1+\sin^{2}2\psi\right),\\
-{\rm i}\sin2\psi\, A &= -2\sin\theta\sin2\psi\cos2\psi-{\rm i}\,\sin2\theta\sin^22\psi,
\end{align*}
so that the real parts cancel identically, the imaginary parts combine, and
$A'-{\rm i}\sin2\psi\,A={\rm i}\sin2\theta$. Hence
$\Xi_{\rm c}'={\rm i}\mathcal K\sin 2\theta\,e^{{\rm i}\phi}$, which is exactly
\eqref{eq:Xi_prime_new}.
The full solution of Eq. \eqref{eq:Xi_prime_new} is therefore
\begin{equation}
\Xi(\tau)=\Xi_{\rm c}(\tau)+B,
\label{eq:A_from_initial}
\end{equation}
with the integration constant $B\in\mathbb C$ fixed by the initial condition,
$B=\Xi(0)-\Xi_{\rm c}(0)$. In real components,
\begin{align}
X_{\rm cm}(\tau)&=X_{O'}+\mathcal K\bigl[\sin2\theta\,\sin2\psi\,\cos\phi
+2\sin\theta\,\cos2\psi\,\sin\phi\bigr],\label{eq:X_closed_form}\\
Y_{\rm cm}(\tau)&=Y_{O'}+\mathcal K\bigl[\sin2\theta\,\sin2\psi\,\sin\phi
-2\sin\theta\,\cos2\psi\,\cos\phi\bigr],
\label{eq:Y_closed_form}
\end{align}
where $(X_{O'},Y_{O'})=(\operatorname{Re}B,\operatorname{Im}B)$ are the coordinates of the
point $O'$ about which the rosette is centred. Thus $\Xi_{\rm c}=\Xi-B$ is the trajectory
referred to $O'$, the position of $O'$ is obtained \emph{exactly}, without any averaging, from the initial conditions: for a
trajectory starting at the origin, $\Xi(0)=0$, one has $B=-\Xi_{\rm c}(0)$, which for
$\psi_0=\pi/4$ and $\theta_0=\sin^{-1}\sqrt{C} < \pi/2$ reduces to $B=-2\mathcal K\sqrt{C(1-C)}\,e^{{\rm i}\phi_0}$.
 
Taking the modulus of \eqref{eq:Xic_closed_form} and using the first integral
\eqref{eq:1intl0}, $C=\sin^{2}\theta\sin 2\psi$, the distance of the centre of mass from
$O'$ becomes a function of the nutation angle $\theta$ alone,
\begin{equation}
\rho(\tau):=\bigl|\Xi_{\rm c}(\tau)\bigr|
=|\mathcal K|\sqrt{\sin^{2}2\theta\,\sin^{2}2\psi+4\sin^{2}\theta\,\cos^{2}2\psi}
=2|\mathcal K|\sqrt{\sin^{2}\theta(\tau)-C^{2}}.
\label{eq:rho_closed_form}
\end{equation}
Since $\theta(\tau+T/2)=\pi-\theta(\tau)$, $\rho$ is $T/2$-periodic. Along any closed
orbit with $0<C<1$ one has $\sin^{2}\theta(\tau)\in[C,1]$, so the minimum and the maximum of $\rho$ follow
immediately,
\begin{equation}
\frac{\rho_{\rm in}}{|\mathcal K|}=2\sqrt{C(1-C)},
\qquad
\frac{\rho_{\rm out}}{|\mathcal K|}=2\sqrt{1-C^{2}} ,
\label{eq:rho_in_out_new}
\end{equation}
attained respectively at $\sin^{2}\theta=C$, i.e.\ $\psi=\pi/4$, and at $\theta=\pi/2$, i.e., $\sin 2\psi=C$. By the
quarter-period symmetry of the $\psi-\theta$ trajectory,  these two instants alternate, separated by exactly $T/4$, so that each
quarter-period carries one passage from the outer to the inner envelope, or from the inner to the outer envelope. The $C$-dependence of the inner and outer envelope radii given by Eq.~\eqref{eq:rho_in_out_new} is plotted in figure \ref{fig:rho_in_out_vs_C}(a). The limiting values are $\rho_{\rm out}/|\mathcal K|\to2$,  and
$\rho_{\rm in}/|\mathcal K|\to0$ as $C\to0^{+}$, and both radii vanish as $C\to1^{-}$.
The annulus radius $\rho_{\rm in}$ is the largest, with $\rho_{\rm in}=|\mathcal K|$, at $C=1/2$. Panel~\ref{fig:rho_in_out_vs_C}(b) illustrates a representative laboratory-frame trajectory for \(C=0.74\); the dashed circles of radii \(\rho_{\rm in}\) and \(\rho_{\rm out}\) visualize the inner and outer envelopes of the horizontal centre-of-mass motion. For the initial value of the spin angle $\psi_0=\pi/4$, the distance $|B|$ from initial centre-of-mass position to $O'$ equals to $\rho_{\rm in}$, as illustrated in figures \ref{fig:xy_pane_mot} and \ref{fig:rho_in_out_vs_C}(b).

\subsubsection*{Radial turning points and origin of cusps}

The geometric meaning of the two radii $\rho_{\rm in}$ and $\rho_{\rm out}$ and the origin of the sharp tips of the rosette will now be explained in the laboratory frame. 

Differentiating equation \eqref{eq:rho_closed_form} for $\rho(\tau)$  and using $\theta'=-\cos2\psi \sin\theta$ we obtain the radial velocity in the closed form,
\begin{equation}
\rho'(\tau)
=
\frac{-2|\mathcal{K}|^2}{\rho(\tau)} \sin \theta(\tau) \,\sin 2\theta(\tau)\,
\cos 2\psi(\tau).
\label{eq:rhosq_prime_new}
\end{equation}

\begin{figure}
    \centering
    \includegraphics[width=0.95\linewidth]{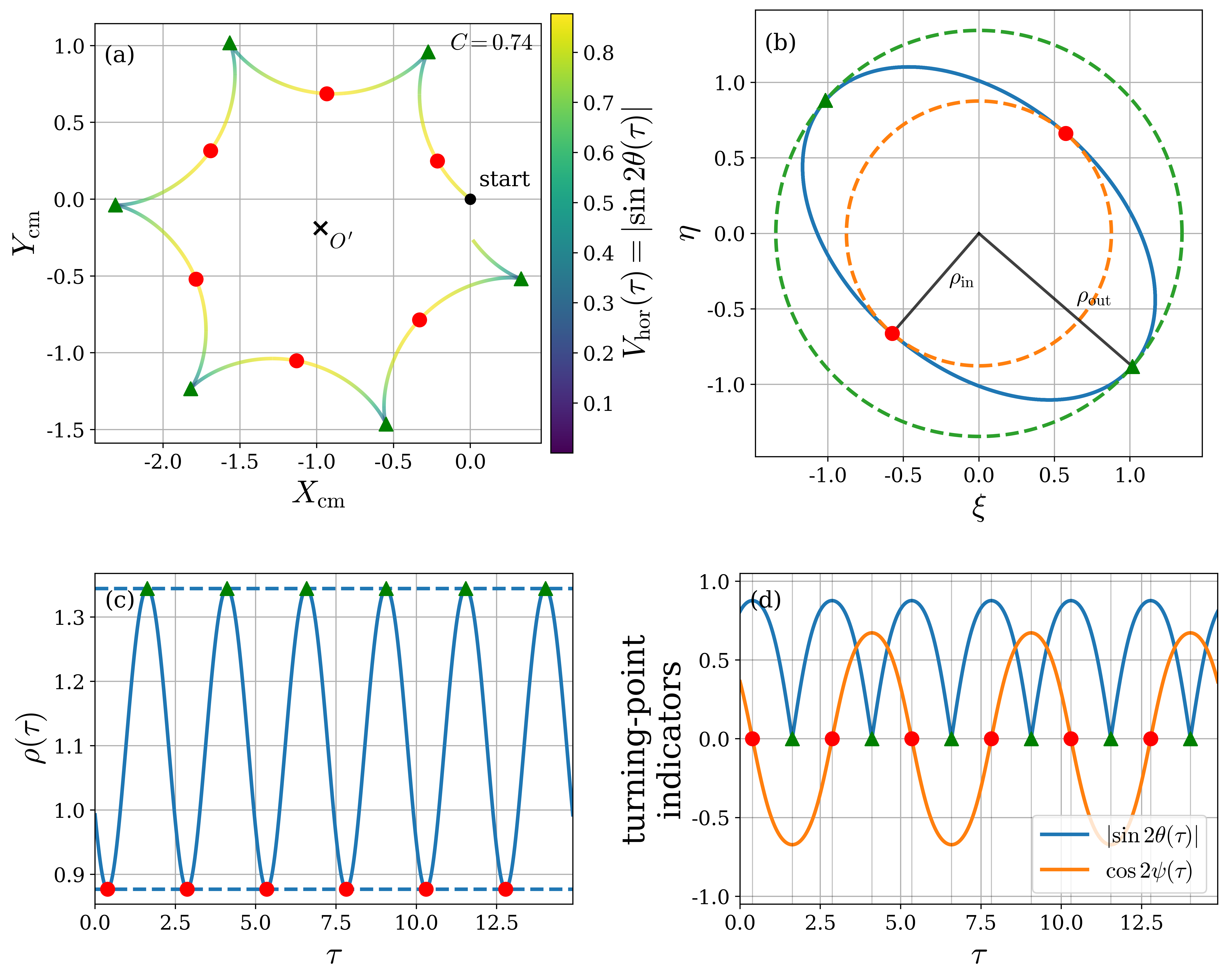}
    \vspace{-0.4cm}
\caption{Turning points and cusps of the rosette (\(C=0.74\)).
(a) Laboratory-frame trajectory \((X_{\rm cm},Y_{\rm cm})\), colour-coded by the instantaneous
horizontal speed \(V_{\rm hor}(\tau)=|\mathcal{K}\sin 2\theta(\tau)|\) (here \(\mathcal{K}=1\)).
(b) Horizontal projection of the centre-of-mass orbit with the centre-of-mass coordinates \((\xi,\eta)\) in the frame co-rotating around $O'$.  Dashed circles mark the
envelope radii \(\rho_{\rm in}\) and \(\rho_{\rm out}\).
(c) Radial distance \(\rho(\tau)\),
shown over \(3T\).
(d) Turning-point indicators \(|\sin 2\theta(\tau)|\) and
\(\cos2\psi(\tau)\).
Red circles mark \(\cos2\psi=0\); green triangles mark \(\sin 2\theta=0\). The initial values of the angles are \(\phi_0=0.9\), \(\psi_0=0.6\), and \(\theta_0=1.1\).}
    \label{fig:turningpoints_cusps}
\end{figure}

Therefore for $0\!<\!C\!<\!1$ the radial turning points, i.e.,  the time instants $\tau=\tau_{*}$ corresponding to the extrema of \(\rho(\tau)\), satisfy
\begin{equation}
\rho'(\tau_*)=0
\quad\Longleftrightarrow\quad
\sin 2\theta(\tau_*)=0
\ \ \text{or}\ \ 
\cos2\psi(\tau_*)=0,
\label{eq:turning_points_new}
\end{equation}
that is, $\theta(\tau_*)=\pi/2$ or $\psi(\tau_*)=\pi/4$ in the fundamental pocket.

These two possibilities have a simple geometric meaning. The first one corresponds to a
vanishing horizontal velocity amplitude, $V_{\rm hor}=0$, since the planar speed is proportional to
\(|\sin 2\theta(\tau)|\). The second means that the instantaneous heading direction in \eqref{eq:heading} becomes perpendicular to the radial centre-of-mass position relative to $O'$, specified in \eqref{eq:X_closed_form}-\eqref{eq:Y_closed_form}.
The time instants $\tau_*$ are locked to quarter-period
symmetry points of the $\psi$-$\theta$ orbit. Therefore, 
the instants corresponding to $\theta=\pi/2$ 
and $\psi=\pi/4$ alternate, separated by exactly $T/4$.  They yield, respectively, $\rho=\rho_{\rm out}$ and $\rho=\rho_{\rm in}$, with
$\rho_{\rm in}<\rho_{\rm out}$ whenever $0<C<1$. From \eqref{eq:rho_in_out_new} it follows that for $C=1$, the horizontal trajectory degenerates to a point, $\rho(\tau)=\rho_{\rm in}=\rho_{\rm out}=0$. The only other degenerate case in which the horizontal motion collapses to a point corresponds to \(\mathcal{K}=0\) (equivalently \(\mu_3=\mu_1\)).

At $\theta(\tau_*)=\pi/2$ the planar speed
$V_{\rm hor}=|\mathcal K\sin2\theta|$ vanishes \emph{exactly} while $\rho$ attains its maximum $\rho_{\rm out}$. 
The approach to $\rho_{\rm out}$ corresponds to the increase of $\rho$, and therefore, owing to equation \eqref{eq:rhosq_prime_new}, to $\theta >\pi/2$. 
Using equations \eqref{eq:Vhor_again}, \eqref{eq:heading},  \eqref{eq:X_closed_form}, \eqref{eq:Y_closed_form}, and \eqref{eq:rho_closed_form},
we compare the direction of the centre-of-mass horizontal velocity to the centre-of-mass horizontal position with respect to $O'$,
\beq
\frac{1}{V_{\rm hor}\, \rho} (X'_{\rm cm},Y'_{\rm cm})\cdot \left(\begin{array}{c}
X_{\rm cm}-X_{O'}\\
Y_{\rm cm}-Y_{O'}
\end{array}\right) \xrightarrow{\theta \rightarrow \pi/2}  \left\{ \begin{array}{c}
1 \mbox{ for } \theta > \pi/2,\\
-1 \mbox{ for } \theta < \pi/2.
\end{array}
\right. \label{coinc}
\eeq 
Therefore, in the limit of $\theta \rightarrow \pi/2$, the trajectory reaches the outer envelope radially (with respect to $O'$), 
and retraces radially inward - the geometry of a cusp. The cusps of the rosette therefore coincide, for every $C\in(0,1)$, with the \emph{outer} envelope $\rho_{\rm out}$. 
As illustrated in figure \ref{fig:xy_pane_mot}, their sharpness increases as $C\to0^{+}$, because $\rho_{\rm out}$ increases, and $\rho_{\rm in}$ decreases to zero, as shown in figure \ref{fig:rho_in_out_vs_C}(a), while the ratio $T/\tau_{\rm drift}$ is almost constant (see the bounds in \eqref{ratio}).

At the complementary instants $\psi(\tau_*)=\pi/4$, $\rho(\tau_*)=\rho_{\rm in}$, the speed remains finite, and the motion turns from inward to
outward without becoming stationary, so the inner envelope is attained smoothly. 

Figure~\ref{fig:turningpoints_cusps} illustrates this mechanism for the representative case
\(C=0.74\). Panel~(a) shows the laboratory-frame trajectory in the \(XY\) plane, colour-coded by the instantaneous horizontal speed. Panel~(b) anticipates the next paragraph and shows the same orbit in the frame rotating
with the mean azimuthal drift; the dashed circles of radii
\(\rho_{\rm in}\) and \(\rho_{\rm out}\) are the envelopes obtained
above and are common to both frames, since the two differ by a pure rotation about $O'$. Panel~(c) displays the periodic radial distance
\(\rho(\tau)\), whose extrema determine the envelope radii. Finally, panel~(d) shows the two factors entering the turning-point condition \eqref{eq:turning_points_new}, namely \(|\sin 2\theta(\tau)|\) and \(\cos2\psi\). Their zeros interleave, separated by $T/4$: the zeros of $|\sin2\theta|$ (green triangles) mark the cusps and the outer envelope, those of $\cos2\psi$ (red circles) the smooth inner turning points.

The horizontal centre-of-mass projections computed by
\citet{joshi} (their Fig.~7) and \citet{vaquero}
(their Fig.~8) for general $C_{2v}$ (but not $S_4$) bodies exhibit smooth,
petal-like loops without sharp tips. By contrast, the rosettes
derived here for $S_4 \cap C_{2v}$ particles generically
display cusps, whose origin is traced in equation \eqref{coinc}. The cusps form when a radial extremum
of $\rho(\tau)$ coincides with the vanishing of the
horizontal speed $|\sin 2\theta(\tau)|$, and they become
progressively sharper as $C \to 0^+$. This coincidence is
facilitated by the $S_4$-enforced equality of the two
transverse translational mobilities, which causes the
horizontal speed to depend on $\theta$ alone
(Eq.~\ref{eq:Vhor_again}). For a general $C_{2v}$ body with
$\mu_1 \neq \mu_2$, the horizontal velocity acquires an
additional dependence on $\psi$ that generically prevents the
simultaneous satisfaction of both conditions, smoothing the
tips into the rounded petals observed in those studies. The
cusps are therefore a distinctive geometric signature of the
$S_4$ subclass and could serve as an experimentally testable
prediction: a particle whose horizontal trajectory shows
sharp tips rather than smooth petals would be identified as
possessing $S_4$ symmetry of its mobility tensor, in addition to $C_{2v}$.

\subsubsection*{The co-rotating frame and the
symmetry of the  orbit centred at $O'$}

It is instructive to exhibit the periodic part of the motion  explicitly in the frame of reference rotating with the angular velocity $\omega_Z$ around $O'$. Writing $\phi=\omega_Z\tau+\phi_{\rm per}$ in \eqref{eq:Xic_closed_form}, we express the horizontal components of centre-of-mass position as
\begin{equation}
W_{\rm per}(\tau):=\Xi_{\rm c}(\tau)\,e^{-{\rm i}\omega_Z\tau}
=\mathcal K\,A(\tau)\,e^{{\rm i}\phi_{\rm per}(\tau)}=:\xi(\tau)+{\rm i}\eta(\tau),
\label{eq:Wc_def}
\end{equation}
which is manifestly $T$-periodic, since $A$ is built from $T$-periodic functions $\theta$ and $\psi$ and $\phi_{\rm per}$ is $T/2$-periodic. Equivalently,
$\Xi_{\rm c}(\tau)=W_{\rm per}(\tau)\,e^{+{\rm i}\omega_Z\tau}.$

Thus $(\xi,\eta)$ are the centre-of-mass horizontal coordinates in the frame that rotates about
$O'$ with the angular velocity $\omega_Z$; because multiplication by
$e^{-{\rm i}\omega_Z\tau}$ is a pure rotation, $|W_{\rm per}|=|\Xi_{\rm c}|=\rho$, i.e.\
\eqref{eq:rho_closed_form} gives the radial distance in either frame, which is why the envelope radii obtained in the laboratory frame apply here unchanged. An exemplary centre-of-mass orbit in the frame co-rotating around $O'$  is shown in figure~\ref{fig:turningpoints_cusps}b. To characterise the radial oscillations of the orbit centred at $O'$, it is convenient to write
\begin{equation}
W_{\rm per}(\tau)=\rho(\tau)\,e^{{\rm i}\chi(\tau)},
\label{eq:polar_Wc}
\end{equation}
where \(\chi(\tau)=\arg W_{\rm per}(\tau)\) is the polar angle in the co-rotating frame. From \eqref{eq:Xic_closed_form} and \eqref{eq:Wc_def} $\chi$ is obtained explicitly by
\begin{equation}
    \chi(\tau)=\phi_{\rm per}(\tau)+\arg A(\tau), \qquad \arg A(\tau) =\rm{atan2}(-2\sin\theta \cos 2\psi,\, \sin2\theta \sin2\psi).
    \label{eq:chi_def}
\end{equation}
At the inner turning instants, when \(\cos2\psi=0\) by \eqref{eq:turning_points_new}, $A$ is real and \(\chi = \phi_{\rm per}\) modulo $2\pi$; this is the co-rotating counterpart of the statement, made in the laboratory frame, that the heading direction is perpendicular to the radius. At the outer turning instant, when $\sin 2\theta =0$, $A$ is imaginary, and \(\chi = \phi_{\rm per}+\pi/2\) modulo $2\pi$.

The closed form \eqref{eq:Xic_closed_form} also fixes the symmetry of the centre-of-mass orbit centred at $O'$. An example of such an orbit is shown in figure~\ref{fig:turningpoints_cusps}b. We now choose $\theta(0)=\pi/2$. Since $A$ is built from the orientation angles alone, the relations \eqref{eq:halfperiod}, \eqref{time_reversal} and \eqref{phi-} translate directly into 
\begin{equation}
A(\tau+T/2)=-A(\tau), \qquad A(-\tau)=-A^*(\tau),
\label{eq:A_symmetries}
\end{equation}
and therefore
\begin{equation}
W_{\rm per}(\tau+T/2)=-W_{\rm per}(\tau),
\qquad
W_{\rm per}(-\tau)=-e^{2{\rm i}\phi_{\rm per}(0)}\,W_{\rm per}^*(\tau),
\label{eq:Wc_symmetries}
\end{equation}
where $(\cdot)^*$ denotes complex conjugation.
The first transformation, $\tau \rightarrow \tau + T/2$, leads to
a point reflection through the origin of the $(\xi,\eta)$ plane. The second one, $\tau \rightarrow -\tau$, with $\tau=0$ determined by the condition $\theta(0)=\pi/2$, leads to a mirror reflection about the line at angle $\phi_{\rm per}(0)+\pi/2$; together they generate two orthogonal mirror axes, so that the  
orbit centred at $O'$ possesses the same symmetry group as an ellipse. Evaluating $A$ at the two consecutive turning instants identified above gives
$A(0)=2{\rm i}\sqrt{1-C^{2}}$, since $\sin2\theta=0$, $\sin2\psi=C$ and
$\cos2\psi=-\sqrt{1-C^{2}}$ there, and $A(T/4)\in\mathbb R$, since $\cos2\psi=0$.
Because $\chi=\phi_{\rm per}+\arg A$ and \(\phi_{\rm per}(\tau)-\phi_{\rm per}(0)\) is an odd function of $\tau$,  
and $T/2$-periodic, so that it vanishes at
$T/4$, the semi-major axis, of length $\rho_{\rm out}$, lies at the angle
\begin{equation}
\chi_{\rm major}=\phi_{\rm per}(0)+\frac{\pi}{2}
\label{eq:major_axis_angle}
\end{equation}
to the $\xi$ axis, and the semi-minor axis, of length $\rho_{\rm in}$, at the angle
$\chi_{\rm minor}=\phi_{\rm per}(0)$. If the time origin is placed at $u=u_+$, as in Eq.~\eqref{defi}, then
simply $\phi_{\rm per}(0)=\phi_0$. Their ratio follows from \eqref{eq:rho_in_out_new},
\begin{equation}
\frac{\rho_{\rm in}}{\rho_{\rm out}}=\sqrt{\frac{C}{1+C}},
\label{eq:axis_ratio}
\end{equation}
increasing monotonically from $0$ at $C\to0^{+}$ to $1/\sqrt2$ at $C\to1^{-}$.
 
The orbit shown in figure \ref{fig:turningpoints_cusps}b resembles an ellipse. Below we verify that uncertainty of this approximation decreases with the increase of $C$. Writing $V:=W_{\rm per}e^{-{\rm i} \phi_{\rm per}(0)}$, the relations \eqref{eq:Wc_symmetries} become $V(-\tau)=-V^*(\tau)$ and $V(\tau+T/2)=-V(\tau)$, i.e.\ $\operatorname{Re}V$ is odd and $\operatorname{Im}V$ even, both anti-periodic with half-period $T/2$. Their Fourier representations therefore contain only odd harmonics of $\varpi=2\pi/T$,
\begin{equation}
\operatorname{Re}V=\!\!\sum_{n\ \text{odd}}\!\!a_n\sin (n\varpi\tau),
\qquad
\operatorname{Im}V=\!\!\sum_{n\ \text{odd}}\!\!b_n\cos (n\varpi\tau),
\label{eq:V_fourier}
\end{equation}
and an ellipse corresponds to retaining the first harmonic alone, all higher coefficients vanishing
identically. The leading departure is therefore measured by the relative weight $|b_3/b_1|$ of the third harmonic, which decreases monotonically with the increase of $C$: it equals $0.13$ at $C=0.1$, $0.011$ at
$C=0.74$ and $3.4\times10^{-3}$ at $C=0.9$, the ratio $|a_3/a_1|$ behaving in the same way and taking the values $0.16$, $0.011$ and $3.5\times10^{-3}$, respectively. The orbit thus becomes elliptical in the centre limit $C\to1^{-}$, where the orientation dynamics linearises about the fixed point, and departs increasingly and significantly from an ellipse as the separatrix is approached at $C\rightarrow 0^+$.

\subsubsection*{Strictly periodic rosettes}
Because $A(\tau)$ in \eqref{eq:Xic_closed_form} depends only on $\theta$ and $\psi$, it is
$T$-periodic, while $\phi(\tau+T)=\phi(\tau)+\Delta\phi$. The closed form therefore implies at once
that after one orientation period the whole centred trajectory is merely rotated about $O'$,
\begin{equation}
\Xi_{\rm c}(\tau+T)=e^{{\rm i}\Delta\phi}\,\Xi_{\rm c}(\tau).
\label{eq:Xic_rotation}
\end{equation}
\begin{figure}
    \centering
\includegraphics[width=0.99\linewidth]{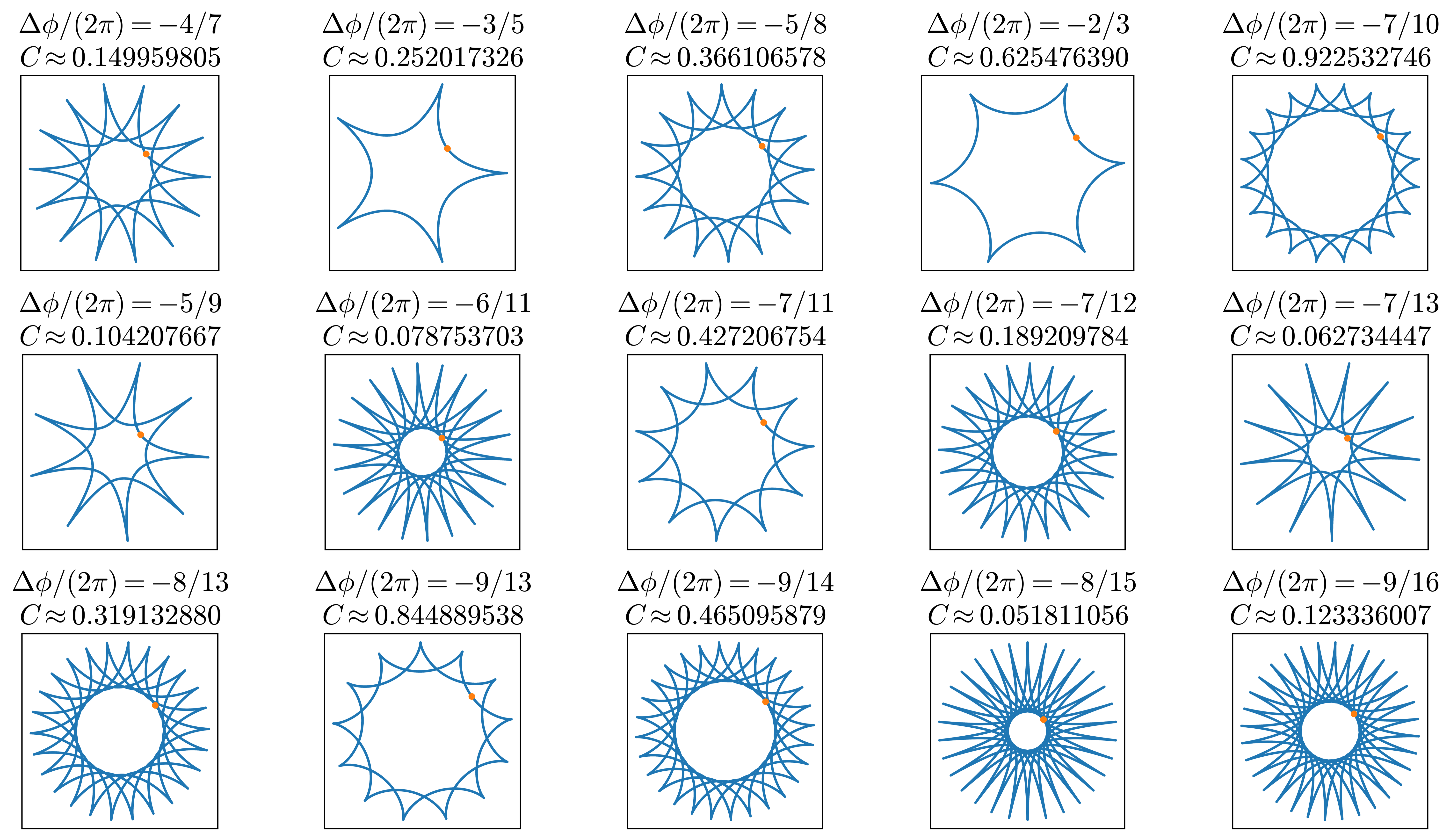}
   \vspace{-0.2cm} \caption{Strictly periodic horizontal projections of the centre-of-mass trajectories (“rosettes”). Each panel corresponds to a rational azimuthal drift $\Delta\phi/(2\pi)=-p/q$ (labelled), with the associated value of the first integral $C$; the curve is shown over a window $2qT$, i.e. twice the orientation closure time $qT$, to highlight repeatability of the pattern. The characteristic rosette envelope radii, $\rho_{\rm out}$ and $\rho_{\rm in}$ , are summarised in Table~\ref{tab:periodic_rosettes}. The initial values of the angles are \(\phi_0=\pi/5\), \(\psi_0=\pi/4\), and \(\theta_0=\sin^{-1}\sqrt{C}<\pi/2\).}
    \label{fig:str_per_ros}
\end{figure}
\begin{table}
\label{eq:Xic_half_r}
\caption{Examples of strictly periodic rosette trajectories in the $XY$ plane, identified by rational values of the azimuthal drift per orientation period, $\Delta\phi/(2\pi)=-p/q$. For a given ratio $p/q$ the planar trajectory closes after $q$ orientation periods, i.e.\ over $qT$. The values of $C$ were obtained numerically by solving $\Delta\phi(C)/(2\pi)=-p/q$. The corresponding inner and outer envelope radii follow from the closed-form
expressions \eqref{eq:rho_in_out_new} and are reported in units of $|\mathcal{K}|$. The orientation
always repeats after $qT$; the planar curve closes after $mT/2$ with $m$ given by \eqref{eq:closure_m}, which for $p,q$ both odd is only half the orientation closure time.}
\centering
\begin{tabular}{cccccc}
\hline
$-\Delta\phi/(2\pi)$ &  $C\ (\approx)$ & orientation
& curve & $\rho_{\rm in}/|\mathcal{K}|$ & $\rho_{\rm out}/|\mathcal{K}|$ \\
 & & closure & closure & & \\ \hline
$4/7$  & $0.149959805$ & $7T$  & $7T$  & $0.7141$ & $1.9774$ \\
$3/5$  & $0.252017326$ & $5T$  &  $5T/2$  & $0.8683$ & $1.9354$ \\
$5/8$  & $0.366106578$ & $8T$  & $8T$  & $0.9635$ & $1.8611$ \\
$2/3$  & $0.625476390$ & $3T$  & $3T$  & $0.9680$ & $1.5605$ \\
$7/10$ & $0.922532746$ & $10T$ & $10T$ & $0.5347$ & $0.7718$ \\
$5/9$  & $0.104207667$ & $9T$  &$9T/2$ & $0.6111$ & $1.9891$ \\
$6/11$ & $0.078753703$ & $11T$ & $11T$ & $0.5387$ & $1.9938$ \\
$7/11$ & $0.427206754$ & $11T$ & $11T/2$ & $0.9893$ & $1.8083$ \\
$7/12$ & $0.189209784$ & $12T$ & $12T$ & $0.7834$ & $1.9639$ \\
$7/13$ & $0.062734447$ & $13T$ & $13T/2$ & $0.4850$ & $1.9961$ \\
$8/13$ & $0.319132880$ & $13T$ & $13T$ & $0.9323$ & $1.8954$ \\
$9/13$ & $0.844889538$ & $13T$ & $13T/2$ & $0.7240$ & $1.0699$ \\
$9/14$ & $0.465095879$ & $14T$ & $14T$ & $0.9976$& $1.7705$ \\
$8/15$ & $0.051811056$ & $15T$ & $15T$ & $0.4433$ & $1.9973$ \\
$9/16$ & $0.123336007$ & $16T$ & $16T$ & $0.6576$ & $1.9847$ 
\end{tabular}
\label{tab:periodic_rosettes}
\end{table}

Moreover, $A(\tau + T/2)=-A(\tau)$ by \eqref{eq:A_symmetries}, while the azimuth advances by exactly half its full-period increment, $\phi(\tau + T/2) = \phi(\tau) + \Delta\phi/2$ by \eqref{phi_half_T}. 
Hence the sharper half-period relation
\begin{equation}
\Xi_{\rm c}\!\left(\tau+\tfrac{T}{2}\right)=e^{{\rm i}(\Delta\phi/2+\pi)}\,\Xi_{\rm c}(\tau).\label{rotation}
\end{equation}
The planar trajectory becomes strictly periodic when the azimuthal drift per orientation period is
commensurate with $2\pi$, i.e.
\begin{equation}
\frac{\Delta\phi(C)}{2\pi}=-\frac{p}{q}\in\mathbb{Q},
\qquad
\Delta\phi(C):=\phi(\tau+T)-\phi(\tau),
\label{eq:rational_drift_condition}
\end{equation}
with integer $p,q$ chosen to have the greatest common divisor $\gcd(p,q)=1$ (i.e., \(p\) and \(q\) are coprime). In that case the orientation repeats after $qT$. The planar curve, however, depends
only on $\Xi_{\rm c}$ and may close earlier: by \eqref{rotation} it repeats after
$mT/2$, where $m$ is the smallest positive integer with
$m(\Delta\phi/2+\pi)\in2\pi\mathbb Z$. Since $\gcd(q-p,q)=\gcd(p,q)=1$ one has
$\gcd(q-p,2q)=\gcd(q-p,2)$ and therefore
\begin{equation}
m=\frac{2q}{\gcd(q-p,2)}=
\begin{cases} q, & p \text{ and } q \text{ both odd},\\[2pt] 2q, & \text{otherwise,}\end{cases}
\label{eq:closure_m}
\end{equation}
so that the rosette closes after $qT/2$ when $p$ and $q$ are both odd, and after $qT$
in all other cases. In the former case the curve closes although the orientation has not yet
returned to its initial value. The values of $C$ for selected rational ratios were obtained
\emph{numerically} by solving $\Delta\phi(C)/(2\pi)=-p/q$ using the closed-form expression
\eqref{eq:delta_phi} for $\Delta\phi(C)$, together with a root-finding
procedure in $C$ with a numerical tolerance of $10^{-12}$.

Examples of strictly periodic rosette trajectories are listed in Table~\ref{tab:periodic_rosettes}, including the rational drift ratio $-\Delta\phi/(2\pi)$, the numerically determined value of $C$, the closure time, and the normalised envelope radii $\rho_{\rm in}/|\mathcal{K}|$ and $\rho_{\rm out}/|\mathcal{K}|$. Representative rosette patterns for selected rational drifts $-p/q$ are shown in Fig.~\ref{fig:str_per_ros}; they were obtained by evaluating the closed form \eqref{eq:Xic_closed_form}
along the exact Euler angles, which avoids the accumulation of integration error over the long closure windows required at large $q$.

The strictly periodic rosettes correspond to a special case. Most of the horizontal projections of the centre-of-mass trajectories in the laboratory frame never close. On the contrary, centre-of-mass trajectories in the co-rotating frame of course are always closed.

\section{Conclusions}\label{sec:conclusions}
A recent trend in the literature is to study dynamics of small rigid particles of different shapes and density distribution, settling under gravity in a viscous fluid, in the low-Reynolds-number systems \citep{tozzi2011,CandelierMehlig2016,Palusa2018,candelier_torques_2025,vaquero,vaquero2025fluttering,joshi,flapper2025settling,Huseby2025,Huseby2026Bifurcations,cheng2026,gissinger2026} and for a small inertia regime \citep{DabadeMarathSubramanian2015,ravichandran2023,candelier_torques_2025,Sundberg2026FluidInertia}. The growing interest is related to phenomena observed in nature, such as sedimenting river flocks \citep{spencer2022,gu2025sedimentation} or pollution spreading in the atmosphere \citep{tatsii}, and also potential applications, such as construction of efficient microrobots \citep{Greenidge2025OloidRobot,flapper2025settling}.

Motivated by these developments, the present paper focuses on analytical derivation of the orientational and translational Stokesian dynamics for a certain class of shapes of 
sedimenting rigid bodies: particles possessing both $C_{2v}$
symmetry (under mirror reflections in two orthogonal planes, $xz$ and $yz$, crossing at $x=y=0$) and the roto-reflection symmetry $S_4$ (rotation by $\pi/2$ about the principal axis $x=y=0$, composed with reflection in the plane $xy$).
This additional $S_4$ constraint is the key to exact solvability of the dynamics. Physically, it forces the $3\times 3$ translational-translational part of the mobility tensor, evaluated with respect to the center of mass in the body reference frame
to be diagonal with $\mu_1 = \mu_2$ for the $x$ and $y$ body-frame components, and it constrains the $3\times3$ rotation-translation coupling part of the mobility tensor in the body reference frame to be symmetric, with a single independent off-diagonal $xy$ parameter~$\mu_b$. Therefore, for the class of particles considered in this paper, the centre of mass is the same as the mobility centre, defined, e.g., by \cite{KimKarrila2005}. Such particles have been called 'cocentred' by \cite{Huseby2025,Huseby2026Bifurcations} who extensively discussed how the shift between both centres affects basic features of the dynamics.

The choice of shapes considered in this paper was motivated by recent studies 
of orientational and translational dynamics of particles with $C_{2v}$ symmetries, performed by \cite{miara2024dynamics,vaquero,vaquero2025fluttering,joshi}. Particles with two orthogonal symmetry planes tend to a stable orientation along gravity (settlers), approach a certain inclined orientation (drifters), or move quasi-periodically (flutterers), depending on the signs of two rotational-translational mobility coefficients, as classified by \cite{joshi}. Examples of the flutterer dynamics were provided experimentally and numerically by \cite{miara2024dynamics,vaquero}. However, for particles with $C_{2v}$ symmetry, there are no exact solutions of their orientational and translational dynamics.  The additional $S_4$ symmetry is essential to obtain the exact solution.

Accordingly, the exact results derived here apply to the
$S_4 \cap C_{2v}$ symmetry shapes and are a subclass of the flutterers. Whether a specific geometry
(e.g., a U-shaped disk of a given curvature) satisfies this
symmetry depends on its $6 \times 6$ mobility tensor (or its inverse, the resistance tensor), which must be
computed separately, e.g., via finite-element \citep{vaquero},  boundary integral
\citep{joshi}, or multipole methods applied to the bead or Stokeslet models of the particle shape \citep{ekiel2009hydrodynamic,Shekhar2026JFM,cheng2026,gissinger2026}. The price of the additional
symmetry $S_4$ is a narrower class of applicable shapes; the payoff
is complete analytical tractability.

To obtain the exact solution of the orientational dynamics, we use the Euler angles. As in the general case \citep{gonzalez}, the coupled first-order ordinary differential equations for the tilt angle of the symmetry axis and the spin angle about it separate. We found their solution in terms of
Jacobi elliptic functions and incomplete elliptic integrals. Each orbit is labeled by a conserved quantity $C$ (in analogy to the constant of motion found by \cite{joshi}).

The azimuthal angle about the vertical axis is expressed through a standard higher elliptic integral. From these closed-form solutions for the orientation angles we further obtain, again analytically, the vertical displacement of the centre of mass as a function of time and, as a particularly compact by-product, the orbit-averaged sedimentation velocity. After completing the analysis presented in Sec.~\ref{sec:exact_euler} we found  that our resulting exact expressions for the Euler angles are equivalent
to that provided by \citet{MakinoDoi2003} for a propeller with two perpendicular blades. In this paper, we provide a more detailed, coherent and comprehensive  description of the dynamics, including plots of the exact solutions for exemplary initial conditions. Also, a comparison with different Jacobi-function representation of \citet{MakinoDoi2003} provides an additional cross-validation
of our analytical results. Moreover, here  we determine and analyse the new exact solution for the horizontal component of the centre-of-mass motion.

The horizontal component of the centre-of-mass motion, which generically traces rosette-like curves (in general
open spirographs) in the laboratory frame, is obtained exactly as well.  The equation governing the time dependence of the  horizontal coordinates  admits an exact particular solution that is algebraic in the orientation angles, so that the horizontal position follows from the Euler angles already determined.
 Three results then follow analytically. First, the centre of the rosette is determined exactly by the initial conditions.  Second, the distance from that centre depends on the  azimuthal angle alone, whence the inner and outer envelope
radii of the horizontal centre-of-mass trajectory are elementary algebraic functions only of the first integral $C$. Third, differentiating the closed form yields
a turning-point condition that identifies the geometric mechanism responsible for the sharp cusps of the
rosette and shows that the two radial extrema are locked to instants separated by exactly a quarter of the
orientation period. In a frame co-rotating with the mean azimuthal drift the same closed form is manifestly
periodic and exhibits two orthogonal mirror axes, whose orientation is set by a single phase; the orbit is
close to, but distinct from, an ellipse. Finally, we catalogue a family of strictly periodic rosettes, selected
by a commensurability condition between the orientation period and the azimuthal drift. The outer and inner radii of the horizontal rosette are proportional to $\mathcal{K}\!=\!\tfrac{\mu_3 - \mu_1}{2\mu_b}$. Therefore, the size of the rosette increases to infinity when $\mu_b$ decreases to zero.

Summarizing, in this paper we present the detailed theoretical description of the dynamics of all the particle shapes with the $S_4 \cap C_{2v}$ symmetry. Our results provide an analytical benchmark that applies to any rigid body in the $S_4 \cap C_{2v}$ class,  once its three independent mobility coefficients, $\mu_1,\;\mu_3$ and $\mu_b$, are known, regardless of the numerical method used to compute them. All analytical
predictions are cross-validated against high-accuracy numerical integration of the full equations of motion,
with agreement at the level of the numerical tolerance. The analytical benchmark presented in this paper
can be used in further numerical and experimental studies of the Stokesian dynamics of various shapes possessing this symmetry, such as oloids \citep{flapper2025settling} or non-chiral propellers made of two circular disks located in perpendicular planes \citep{MakinoDoi2003, MakinoDoi2005, KimKarrila2005}. Moreover, our exact solution applies also for a non-uniform density distribution satisfying the same $S_4 \cap C_{2v}$ symmetry. It also provides a natural zeroth-order state for future perturbative extensions to lower-symmetry shapes or density distributions. 

From a practical perspective, the closed-form solutions offer several advantages over direct numerical time integration.
First, they serve as high-accuracy benchmarks: the maximum
absolute deviation between the exact and numerical (DOP853)
solutions is of order $10^{-11}$ over the intervals plotted
in figures~\ref{fig:angles} and \ref{fig:cm_motion}, validating both the analytical derivation and the numerical implementation. Second, the explicit dependence of all physical quantities on $C$ enables rapid parametric surveys over the full space of initial
conditions without time-stepping, which is relevant for initially random orientations. Third, the Fourier-based construction of the co-rotating trajectory (Eq.~\ref{eq:V_fourier}) provides a spectral characterisation of the horizontal motion that would be difficult to extract from long-time laboratory-frame integrations alone.

Several directions for future work emerge naturally from the
present analysis. First,  a natural strategy is to treat the departure from $S_4\cap C_{2v}$ symmetry as a small parameter and develop a perturbation theory around the exact solution derived here, using the known Jacobi-function and elliptic-integral
expressions as the zeroth-order approximation. Such an
approach would yield explicit corrections to all the expressions derived in this work.  Second, the inclusion of weak inertial corrections - at the level of the Oseen approximation
\citep{KhayatCox1989} - would perturb the conserved
quantity $C$ itself, causing a slow drift between orbits and
potentially selecting a preferred long-time orientation, as
occurs for sedimenting spheroids
\citep{DabadeMarathSubramanian2015}, slender fibres
\citep{SubramanianKoch2005,Roy2019inertial}, and asymmetric
dumbbells \citep{CandelierMehlig2016}. Here again, the exact
solution provides a natural zeroth-order state for a
systematic expansion in the small Reynolds number, with the
adiabatic evolution of $C(\tau)$ governed by the inertial
torque averaged over one orientation cycle. Third, for sub-micron
particles, Brownian rotational diffusion would compete with
the deterministic orientational dynamics derived here; the
resulting Smoluchowski equation \citep{Risken1989,Dhont1996}
on the $(\psi, \theta)$-torus, with a known deterministic
drift and the first integral $C$ providing the natural
coordinates, could potentially be analysed by spectral methods
that exploit the elliptic structure of the drift term,
extending the classical framework of \citet{Brenner1974} to
particles with $S_4$ roto-reflection symmetry. Next, the exact solution derived here could be helpful for designing efficient microrobots, such as the oloid one studed by \cite{Greenidge2025OloidRobot}, with their motion controlled by a desired choice of the initial orientations.

\backsection[Funding]{This work was supported in part by the National Science Centre under grant UMO-2021/41/B/ST8/04474.}

\backsection[Declaration of interests]{The authors report no conflict of interest.}

\backsection[Author ORCIDs]{Piotr Zdybel, \href{https://orcid.org/0000-0001-7484-1425}{https://orcid.org/0000-0001-7484-1425}, Maria L. Ekiel-Je\.zewska, \href{https://orcid.org/0000-0003-3134-460X}{https://orcid.org/0000-0003-3134-460X}}

\appendix
 
% ---------------------------------------------------------------------
\section{Closed-form solution for $\phi$}
\label{app:phi_derivation}
 
Equation~\eqref{eq:phipl02} gives
\begin{equation}
\phi(\tau)-\phi_0=-2\int_{0}^{\tau}\frac{u(\tau')}{[u(\tau')]^2+1}\,\mathrm{d}\tau'
=-\frac{2}{\Omega}\int_{\zeta_0}^{\zeta}\frac{u(\zeta')}{[u(\zeta')]^2+1}\,\mathrm{d}\zeta',
\label{eq:phi_int}
\end{equation}
with $\zeta=\Omega(\tau-\tau_0)$. We decompose the integrand into partial fractions,
$\tfrac{u}{1+u^2}=\tfrac12\bigl(\tfrac{1}{u-i}+\tfrac{1}{u+i}\bigr)$, and insert the Jacobi
solution $u=u_-+\delta\,\mathrm{cn}^2(\zeta,k)$ of Eq.~\eqref{sol_eq}, which gives
\begin{equation}
u\mp i=(u_-\mp i)\bigl[1-\tilde n_\pm\,\mathrm{cn}^2(\zeta,k)\bigr],
\qquad
\tilde n_\pm=-\frac{\delta}{u_-\mp i},
\qquad
\tilde n_-=\tilde n_+^{*}.
\label{eq:app_partial}
\end{equation}
The standard reduction formula \citep{Lawden1989Elliptic}, understood for complex $\tilde n$
by analytic continuation,
\begin{equation}
\int_{\zeta_0}^{\zeta}\frac{\mathrm{d}\zeta'}{1-\tilde n\,\mathrm{cn}^{2}(\zeta',k)}
=\frac{1}{1-\tilde n}\Bigl[\Pi\bigl(n;\mathrm{am}(\zeta,k^2)\big|k^{2}\bigr)
-\Pi\bigl(n;\mathrm{am}(\zeta_0,k^2)\big|k^{2}\bigr)\Bigr],
\quad
n:=-\frac{\tilde n}{1-\tilde n},
\label{eq:app_reduction}
\end{equation}
applies to each partial fraction separately. Because
$(u_-\mp i)(1-\tilde n_\pm)=u_-\mp i+\delta=u_+\mp i$, the two prefactors combine into a single
turning-point expression and the corresponding characteristics are
\begin{equation}
n_{\pm}=\frac{\delta}{u_+\mp i}=\delta\,\frac{u_+\pm i}{1+u_+^{2}},
\label{eq:app_char}
\end{equation}
so that $n_+$ is precisely Eq.~\eqref{eq:el_char} and $n_-=n_+^{*}$. Since
$\Pi(n^{*};\varphi|k^2)=\Pi(n;\varphi|k^2)^{*}$ for real $\varphi$ and real $k^2$, the two
partial-fraction contributions are complex conjugates of one another and their sum is real,
\begin{equation}
\int_{\zeta_0}^\zeta\Bigl(\frac{1}{u-i}+\frac{1}{u+i}\Bigr)\mathrm{d}\zeta'
=2\,\mathrm{Re}\Bigl\{\frac{1}{u_+-i}\bigl[\Pi(n_+;\mathrm{am}(\zeta,k^2)|k^2)
-\Pi(n_+;\mathrm{am}(\zeta_0,k^2)|k^2)\bigr]\Bigr\}.
\label{eq:app_sum}
\end{equation}
Inserting \eqref{eq:app_sum} into \eqref{eq:phi_int} and using $n_+/\delta=1/(u_+-i)$ yields the
closed form quoted as Eq.~\eqref{eq:phi_sol}. The branch structure of $\Pi$ for a complex
characteristic, and the increment acquired as $\zeta$ crosses integer multiples of $2K(k^2)$,
are discussed in Sec.~\ref{sec:drift} and in \citet{ArmitageEberlein2006}.
 
% ---------------------------------------------------------------------
\section{Closed-form vertical displacement $Z_{\rm cm}(\tau)$}
\label{app:Zcm_derivation}
 
In Eq.~\eqref{eq:z_vel} it is convenient to replace $\mathrm{cn}$ by $\mathrm{dn}$. Using
$\mathrm{dn}^2(\zeta,k)=1-k^2+k^2\mathrm{cn}^2(\zeta,k)$ together with the parameter relations
$u_-=u_+(1-k^2)$ and $\delta=u_+k^2$ (Eqs.~\eqref{pm} and \eqref{defi}), the Jacobi solution
\eqref{sol_eq} collapses to the compact form
\begin{equation}
u(\tau)=u_+\,\mathrm{dn}^2(\zeta,k),
\qquad \zeta=\Omega(\tau-\tau_0).
\label{eq:app_u_dn}
\end{equation}
With $\mathrm{d}\tau=\mathrm{d}\zeta/\Omega$, Eq.~\eqref{eq:z_vel} then integrates to
\begin{equation}
Z_{\rm cm}(\tau)-Z_{\rm cm}(0)=\frac{\mathcal{K}}{\Omega}\left[-\mathcal{V}_0(\zeta-\zeta_0)
+C\int_{\zeta_0}^\zeta\left(u_+\mathrm{dn}^2(\zeta',k)
+\frac{1}{u_+\,\mathrm{dn}^2(\zeta',k)}\right)\mathrm{d}\zeta'\right].
\label{Z_int}
\end{equation}
Both quadratures reduce to incomplete elliptic integrals under the substitution
$\vartheta=\mathrm{am}(\zeta',k^2)$, for which
$\mathrm{d}\vartheta=\mathrm{dn}(\zeta',k)\,\mathrm{d}\zeta'$ (a direct consequence of
\eqref{am_def}) and $\mathrm{dn}(\zeta',k)=\sqrt{1-k^2\sin^2\vartheta}$
\citep{AbramowitzStegun1964,GradshteynRyzhik2007}. For the first integral this gives, by the
definition \eqref{eq:E_def} of $E(\varphi|k^2)$,
\begin{equation}
\int_{\zeta_0}^\zeta \mathrm{dn}^2(\zeta',k)\,\mathrm{d}\zeta'
=\int_{\mathrm{am}(\zeta_0,k^2)}^{\mathrm{am}(\zeta,k^2)}\sqrt{1-k^2\sin^2\vartheta}\,\mathrm{d}\vartheta
= E(\mathrm{am}(\zeta,k^2)|k^2)-E(\mathrm{am}(\zeta_0,k^2)|k^2).
\label{eq:app_E_reduction}
\end{equation}
For the second, the same substitution supplies $\mathrm{d}\zeta'=\mathrm{d}\vartheta/\mathrm{dn}$,
so that
\begin{equation}
\int\frac{\mathrm{d}\zeta'}{\mathrm{dn}^{2}(\zeta',k)}
=\int\frac{\mathrm{d}\vartheta}{\mathrm{dn}^{3}}
=\int\frac{\mathrm{d}\vartheta}{(1-k^2\sin^2\vartheta)\sqrt{1-k^2\sin^2\vartheta}},
\end{equation}
which is exactly the integrand of $\Pi(n;\varphi|k^2)$ in Eq.~\eqref{eq:Pi} at the characteristic
$n=k^2$; hence
\begin{equation}
\int_{\zeta_0}^\zeta \frac{\mathrm{d}\zeta'}{\mathrm{dn}^2(\zeta',k)}
= \Pi(k^2;\, \mathrm{am}(\zeta, k^2) \,|\, k^2)-\Pi(k^2;\,\mathrm{am}(\zeta_0, k^2) \,|\, k^2).
\label{eq:app_Pi_reduction}
\end{equation}
Substituting \eqref{eq:app_E_reduction} and \eqref{eq:app_Pi_reduction} into \eqref{Z_int} and
distributing the prefactors $u_+$ and $u_+^{-1}$ yields Eq.~\eqref{Zcm}. In contrast to
Appendix~\ref{app:phi_derivation}, the characteristic here is real, $n=k^2\in(0,1)$ for
$0<C<1$, so no analytic continuation in the parameter is required.
 
% ---------------------------------------------------------------------
\section{Orbit-averaged sedimentation velocity $\overline{V}_Z$}
\label{app:Vz_derivation}
 
Since $Z'_{\rm cm}(\tau)$ is $T/2$-periodic (Sec.~\ref{sec:z_vel}), its average 
\eqref{eq:Vz_mean_def_new} may equally be evaluated over a full orientation period $T$,
\begin{equation}
\overline{V}_Z=\frac{1}{T}\int_0^T Z'_{\rm cm}(\tau)\,\mathrm{d}\tau
=\frac{Z_{\rm cm}(\tau+T)-Z_{\rm cm}(\tau)}{T},
\label{eq:app_Vz_mean_def}
\end{equation}
which is convenient because the increment then follows directly from the closed form
\eqref{Zcm}, without a separate quadrature of $\sin^2\theta(\tau)$. Over one period the argument
advances as $\zeta\mapsto\zeta+2K(k^2)$, and $\mathrm{am}(\zeta+2K(k^2),k^2)=\mathrm{am}(\zeta,k^2)+\pi$
(Sec.~\ref{sec:drift}). By the $\pi$-periodicity of $\sin^2\vartheta$ in the integrands
\eqref{eq:E_def} and \eqref{eq:Pi},
\begin{equation}
E(\varphi+\pi\,|\,k^2)=E(\varphi\,|\,k^2)+2E(k^2),
\qquad
\Pi\bigl(k^2;\varphi+\pi\big|k^2\bigr)=\Pi\bigl(k^2;\varphi\big|k^2\bigr)+2\Pi(k^2|k^2),
\label{eq:app_EPi_periodicity}
\end{equation}
the second relation being Eq.~\eqref{eq:legendre_connection_Pi} at $n=k^2$; here $E(k^2)\equiv E(\pi/2|k^2)$ and
$\Pi(k^2|k^2)\equiv\Pi(k^2;\pi/2|k^2)$ denote the complete integrals, and the factor $2$
reflects the symmetry of the integrands about $\vartheta=\pi/2$.
 
Applying \eqref{eq:app_EPi_periodicity} to \eqref{Zcm}, the terms evaluated at
$\mathrm{am}(\zeta_0,k^2)$ cancel in the increment, the linear term advances by
$-\mathcal V_0\cdot 2K(k^2)$, and division by $T=2K(k^2)/\Omega$ removes the common factor
$2K(k^2)$, leaving
\begin{equation}
\overline{V}_Z=\mathcal K\left[-\mathcal V_0
+\frac{C}{K(k^2)}\left(u_+E(k^2)+\frac{1}{u_+}\Pi(k^2|k^2)\right)\right].
\label{eq:mean_vel_intermediate}
\end{equation}
The two complete integrals collapse to one by the standard identity
$\Pi(k^2|k^2)=E(k^2)/(1-k^2)$. Moreover,
$u_+(1-k^2)=u_-=1/u_+$ (Eqs.~\eqref{pm} and \eqref{defi}), so that
$[u_+(1-k^2)]^{-1}=u_+$: the two terms of \eqref{eq:mean_vel_intermediate} contribute equally,
and
\begin{equation}
C\left(u_++\frac{1}{u_+(1-k^2)}\right)=2Cu_+=2\left(1+\sqrt{1-C^2}\right),
\label{eq:app_coefficient_reduction}
\end{equation}
which yields the compact form quoted as Eq.~\eqref{eq:mean_vel}.

% Bibliography generated in advance for arXiv; no BibTeX run is required.

\end{document}